# Modeling and Representing Conceptual Change in the Learning of Successive Theories: The Case of the Classical-Quantum Transition

**Giacomo Zuccarini[1］(ORCID ID: 0000-0002-6690-3419)**

**Massimiliano Malgieri[1] (ORCID ID: 0000-0002-9254-2354)**

[1]Physics Department, University of Pavia, via Agostino Bassi 6, 27100, Pavia, Italy.

**Abstract**

Most educational literature on conceptual change concerns the process by which introductory students acquire scientific knowledge. However, with modern developments in science and technology, the social significance of learning successive theories is steadily increasing, thus opening new areas of interest to discipline-based education research, e.g., quantum logic, quantum information and communication. Here we present an initial proposal for modeling the transition from the understanding of a theory to the understanding of its successor and explore its generative potential by applying it to a concrete case – the classical-quantum transition in physics. In pursue of such task, we make coordinated use of contributions not only from research on conceptual change in education, but also on the history and philosophy of science, on the teaching and learning of quantum mechanics, on mathematics education. By means of analytical instruments developed for characterizing conceptual trajectories at different representational levels, we review empirical literature in the search for the connections between theory change and cognitive demands. The analysis shows a rich landscape of changes and new challenges that are absent in the traditionally considered cases of conceptual change. In order to fully disclose the educational potential of the analysis, we visualize categorical changes by means of dynamic frames, identifying recognizable patterns that answer to students' need of comparability between the older and the new paradigm. Finally, we show how the frame representation can be used to suggest pattern-dependent strategies to promote the understanding of the new content, and may work as a guide to curricular design.

# 1 Introduction

## 1.1 The Conceptual Change Revolution: Special Position of Quantum Mechanics within the Research Area

Conceptual change (CC) has been for decades among the most central areas of research in science education. Its rise represented a major break in the account of cognitive development, moving away from a domain general view with a focus on logical structures to a content-based and domain-specific one (Amin et al., 2014). Despite the differences between CC in the learning sciences, concerning the knowledge system of an individual, and CC in the philosophy or history of science, concerning instead the conceptual structure of scientific theories[1], or the ideas shared by a scientific community, a fundamental contribution came by Thomas Kuhn's work (e.g., 1962). His view of theories as substantive systems (rather than logical ones), that assign meaning to relevant concepts, provided a framework for privileging content in cognitive development (Amin et al., 2014). Thus, it became clear that the goal of education, teaching for understanding, could not be achieved without a characterization of students' initial understanding of content knowledge in the domain at hand (Carey, 2000). On this basis, it was possible to examine the needed changes in content, in related factors (epistemology, metacognition, etc.), and to identify productive learning mechanisms and effective strategies to promote CC.

In the following decades, most research focused on the transition from naïve to scientific knowledge, giving rise to ongoing debates on the nature and features of the former, e.g., as concerns the extent of coherence or fragmentation of naïve ideas (Amin & Levrini, 2017). However, in more recent years, we have seen a steady increase of research on the learning of successive scientific models or theories in the context of formal instruction, primarily in chemistry (for a review on quantum chemistry, see Tsaparlis, 2013) and physics (see, e.g., Levrini, 2014; Krijtenburg-Lewerissa et al., 2017, for reviews on relativity and quantum physics). In addition to the knowledge provided on specific difficulties, conceptual resources and effective educational strategies in these topics, the existing research could be instrumental for characterizing a different type of CC: the processes involved in learning a successor of a scientific paradigm already familiar to students. As a matter of fact, while significant part of these processes may be analogous to those related to the learning of introductory science, it is important to investigate whether there are major differences, and to outline how the features of a specific theory change constrain learning mechanisms.

---

[1] Philosophical investigation may examine the nature of changes in the structure of scientific concepts or theories without focusing primarily on the historical circumstances of their introduction, or the socio-scientific dynamics that produced them. See, e.g., Masterton et al. (2016).

Shifts in theory and practice are currently covered in the curriculum of the most diverse disciplines, e.g., from neoclassical to behavioral microeconomics (Dhami, 2016). However, the transition to a quantum picture of the world represents an extraordinary case, as the effects of this revolution span across multiple knowledge domains. At first, the birth of quantum mechanics (QM) essentially affected two natural sciences: physics and chemistry. Now, a second revolution is unfolding (Monroe et al., 2019; Riedel et al., 2019), leading to theory changes in logic (e.g., from the Boolean lattice to the quantum lattice), computer science (e.g., from classical to quantum bits, gates, and algorithms), information and communication technology (e.g., with the transition to quantum communication protocols), and – in response - to the launch of far-reaching educational programmes in the US (House of Representatives, 2018) and the EU (European Quantum Flagship, 2020). From an educational perspective, the revision of concepts of the older theory in the learning of its quantum successor differs from domain to domain: in chemistry, for instance, it primarily involves specific aspects of the Bohr model and the old quantum theory - atomic structure, electron orbits, models of chemical bonding (Tsaparlis, 2013) - rather than measurement and time evolution of a wide range of classical systems, as in physics. Educational research on the classical-quantum transition in disciplines involved in the second quantum revolution is still at the beginning. A wide new area of interest is opening to discipline-based education research.

**1.2 Purposes of this Study**

We endeavor to develop an initial proposal to model CC in learning *the successor of a theory students are familiar with*, and to explore its generative potential by applying it to a concrete case, with the goal to provide a structural account of the educational transition between the theories under scrutiny, and to suggest fruitful teaching strategies for learning the new paradigm. Since most empirical literature on the learning of a successive theory concerns the transition from classical mechanics (CM) to non-relativistic QM in a physics class, we identify it as an ideal context of development and application of the model. In this process, the key element is represented by a fundamental and distinctive feature of CC to a successive theory: the possibility of tracing the trajectories of relevant concepts in theory change. We endeavor to use it for identifying instruments and techniques apt to interpret existing empirical results, for developing representational tools to support concept building, and for devising trajectory-dependent strategies that facilitate the learning of the new content.

In order to describe these purposes in detail and highlight their significance in the educational debate on the theory change at hand, we need to discuss first the results of research on the learning of QM and related issues.

Today, not only university physics curricula worldwide, but also secondary school curricula in many countries (e.g., Stadermann et al., 2019) include instruction on QM, requiring students to revise existing conceptual structures on CM and classical physics at large. Research on learning difficulties has been recently summarized in comprehensive reviews: see, e.g., Krijtenburg-Lewerissa et al. (2017) as regards secondary school and lower undergraduate education, and Marshman and Singh (2015) at upper undergraduate and graduate level.

In both works, the radical differences between classical and quantum phenomena, and the substantial changes in their theoretical description, are recognized as a central element behind the strong challenges students face in learning QM (see Krijtenburg-Lewerissa et al., 2017, p. 1, and Marshman & Singh, 2015, p. 5). Moreover, Marshman and Singh develop a general framework for understanding the patterns of student difficulties in learning QM. Based on the results of educational research, they make the case for a close correspondence between the reasons for insufficient development of expertise in introductory physics and in QM, therefore resulting in analogous patterns of learning difficulties (inconsistent and/or context-dependent reasoning, lack of transfer, etc.). According to the authors, the patterns arise from the combined effect of diversity in the student population (inadequate prior preparation, lack of clear goals and of sufficient motivation to excel) and the paradigm shift.

These reviews are of great value to researchers and instructors, for providing a broad list of difficulties and corresponding interpretive keys, and for providing an argument in favor of adapting strategies and tools which have proven effective in introductory physics to the learning of QM in upper division courses.

However, their findings and claims open further questions to the research community: according to Singh and Marshman (2015), research shows that students in upper-level QM "have common difficulties independent of their background, teaching style, textbook, and institution" (p. 3), and both reviews identify the change from a classical to a quantum picture as a fundamental source of student difficulties. Taken together, these two claims suggest that the main engine behind the emergence of learning difficulties are the educational implications of specific aspects of theory change. Nevertheless, while the reviews mention some differences between the two theories that might favor the onset of specific difficulties, a global analysis of the connection between theory change and learning challenges elicited by research is currently lacking. Therefore, a first goal of this article it to use the model here presented as a guide in the development of clear and comprehensive pictures of the learning challenges related to theory change from CM to QM (Purpose 1).

Second, the CC from naïve physics to CM involves the transformation of conceptual structures formed under the influence of lay culture. The change from CM to QM, instead, implies the transformation of conceptual structures concerning a scientific theory, and developed in the context of formal instruction. How far does the analogy between challenges faced by students at both levels hold? We show that our model can act as a guide in the exploration of its boundaries (Purpose 2).

Third, building a formal and imaginative bridge between the classical and the quantum domain represents a basic cognitive need that emerges in empirical research on student understanding (Levrini & Fantini, 2013; Ravaioli & Levrini, 2017). Other modern theories such as special relativity present a clear demarcation line with classical phenomena: the value of the velocity, as compared to the speed of light, is "an intelligible and effective criterion for guiding imagination toward abstract spaces defined by formalism" (Levrini & Fantini, 2013, p. 1904). On the contrary, the classical limit of QM is an advanced and controversial issue (Klein, 2012). Based on our implementation of Purpose 1, we develop an instrument to visualize continuity and change in educationally relevant aspects of conceptual trajectories, with the aim to help students draw a comparison of classical and quantum notions (Purpose 3).

Last, most research focuses only on difficulties. There is an ever-growing consensus on the importance to shift away from student misconceptions to conceptual resources, i.e., elements of prior knowledge and intuition that can be productively used in the creation of new knowledge (e.g., Vosniadou & Mason, 2012, Coppola & Krajcik, 2013; Amin et al., 2014). We explore what strategies are suggested by the visual representation for supporting the learning of concepts involved in theory change and, specifically, for a productive use of prior intuition in this process (Purpose 4).

**1.3 Outline**

As illustrated in this section, the construction of the model and the achievement of the purposes require a coordinated use of frameworks, tools and evidence from different fields of research, such as CC in education, philosophy and history of science, the teaching and learning of QM and of mathematics.

In Section 2, we build a map of the possible contributions of different factors of learning to the process by which students acquire the knowledge of a successive theory. Since the initial state of the learner is represented by the knowledge of a scientific theory, by drawing on CC literature and philosophical literature on the characterization of scientific knowledge, we identify two cognitive signatures of this knowledge: the understanding of and the ability to use in description, explanation and problem-solving 1) public representations of relevant concepts, e.g., linguistic, mathematical, diagrammatical, and 2) theory-specific exemplars, including laboratory assignments, exercises, etc.. Theory change is always accompanied by changes in both factors. We also discuss changes in epistemic cognition, which presents a complex landscape and possibly an interplay with other factors.

In Section 3, we further develop and operationalize the model, examining what specific changes may occur in these factors in the context of the classical-quantum transition. While we discuss dynamics affecting public representations and exemplars, and issues concerning epistemology, the main focus is on changes in concepts at three representational levels (Section 3.1). In relation to this factor, we identify the internal aspects of each representation and the nature of the changes affecting them – i.e., *ontological, metaphorical,* and *representational* -, and develop a detailed taxonomy of representational ones. While substantiated with examples taken from the classical-quantum transition, this description is general in nature and might be adapted to examine the transition between other theories that use these types of representation.

Section 4 contains an exploration of the connections between the classical-quantum transition - in the form described in Section 3.1 - and learning challenges reported in literature. After a selection of basic classical and quantum concepts and constructs, we build tables reporting their dynamics in detail (included in the supplementary file titled "Dynamics of Theory Change from CM to QM"). Trajectories are structured according to a simplified version of the account of conceptual dynamics in the history of science given by Arabatzis (2020) and to the taxonomy presented in Section 3.1. An interpretive review of empirical literature follows, in which difficulties in learning QM are classified based on the relevant representational level and the specific conceptual trajectories involved. Such review is instrumental to illustrating our implementation of Purpose 1 and 2.

In Section 5, the focus is restricted to changes in the categorical structure of concepts and constructs. By drawing on the global picture of the connections between conceptual trajectories and learning challenges (Purpose 1), and on instruments that have been used for tracing CC in the history of science, such as dynamic frames (see Andersen et al., 2006), we revise this representational tool and adapt it to our needs. The same selection of concepts and mathematical constructs discussed in Section 4 is represented in terms of dynamic frames, visualizing individual trajectories and highlighting educationally significant elements of continuity and change (Purpose 3). By means of the frame representation, we identify three types of regularity in concept evolution from CM to QM: categorical generalization, value disjunction, change in value constraints.

The detailed representation of conceptual dynamics and the identification of the three patterns allow us to suggest pattern-dependent strategies to revise classical concepts and build a consistent knowledge of their quantum version, leveraging prior intuition according to the pattern involved (Purpose 4). Sections 6 presents three examples, one for each pattern.

**2 Towards a Model of the Transition between Successive Theories in Formal Learning**

Instruction-based CC refers to a type of learning requiring the substantial revision of existing knowledge under conditions of systematic instruction (Vosniadou & Mason, 2012). CC contrasts with less problematic learning, such as skill acquisition and acquisition of facts, and in general cannot be achieved simply by means of knowledge enrichment, a mechanism on which we naturally rely, that involves generalization over perceptual experiences or elaboration of knowledge in terms of existing concepts (Amin, 2017).

After decades of research on the transition between naïve and scientific knowledge, there seems to be a general agreement (see the reviews of Vosniadou & Mason, 2012, diSessa, 2014, and Amin et al., 2014) that CC does not occur suddenly, but requires the gradual elaboration and revision of complex knowledge systems consisting of many interrelated elements. It is a difficult process whose promotion requires the interplay of multiple instructional strategies. In addition, it may involve not only changes in learners' cognition, but also in metacognition, epistemic beliefs, beliefs about learning and other factors (e.g., interest, attitudes). Finally, an extensive sociocultural support is needed for achieving all these kinds of change.

In the light of these facts, it is immediate to see that developing a model of CC in the learning of a successor of a scientific theory is a complex task, as there is a need to clarify whether each of the aforementioned factors is affected in a different way by this type of change. However, since the very idea of CC arose from a new awareness of the importance of students' initial understanding of content knowledge, the starting point for the construction of a model can only be a comparison of the initial state of knowledge of the learner in the two cases.

The first issue to address is the dichotomy between coherence and fragmentation. While all researchers agree on basic characters of naïve knowledge as a form of organization of perceptual experience and information received in the context of lay culture (allowing children to provide explanations, make predictions, etc.), they are divided as regards the coherence of children's knowledge system on specific domains. No one takes the stand that children are completely unsystematic in their thinking about physics or biology (diSessa, 2014). The central question is the specification of the nature and the extent of systematicity. From this depends the relation between pre-instructional knowledge and the scientific knowledge to be acquired (often incompatible, according to the "coherence side of the fence", or not at all, for the "pieces" side) and the relative contribution of different instructional strategies (knowledge revision or knowledge integration). However, in the case of the initial state of students' knowledge when learning a successor of a scientific theory known to them, we can reasonably assume they worked for a long time immersed in the environment of the older theory, with the aim to acquire a functional understanding of the domain, and developing complex knowledge structures in the process. These structures may well be incompatible with those of a successor of the older theory and a process of knowledge revision is hardly avoidable. Therefore, we take a "coherence" stance on the understanding of the older theory.

In order to identify other distinctive aspects of this understanding and the forms of change it may undergo, we contrast naïve knowledge according to the framework theory view (Vosniadou & Skopeliti, 2014), a notable representative of the coherence side of the fence, and the knowledge of a scientific theory. A framework theory is a form of organization of everyday experience within given domains: e.g., biology, mathematics. It has a theoretical nature, since it is a principle-based system with a distinct ontology, distinct mechanisms of causality, it is generative (giving rise to predictions and explanations) and constrains the meaning assigned to relevant notions. However, as explained by the proponents of this model, various features separate the understanding of a framework theory from that of a scientific theory (Table 1).

**Table 1** Contrasting some features of the knowledge of a framework theory with the knowledge of a scientific theory

| Knowledge of a Framework Theory | Knowledge of a Scientific Theory |
|---|---|
| minimal representational capacity, not socially shared | understanding and use of public representations: linguistic, mathematical, diagrammatical, etc. |
| loose internal consistency with low explanatory power | internally consistent, high explanatory power |
| not subject to metaconceptual awareness | metaconceptual awareness of disciplinary content and related problem solving |
| not systematically tested for confirmation or falsification | knowledge of the practices used by the relevant community for confirmation/falsification |

In what follows, we analyze the possible contribution of different factors of learning to the CC in the transition from the knowledge of a theory to the knowledge of its successor. In order to structure an initial proposal for the characterization of this change, the understanding of the older theory at the beginning of instruction on the new one is proxied by an optimal state of knowledge of its concepts and other concurrent factors (see Table 1). We are aware that student understanding of the older theory might be far less than optimal in relation to the prerequisites for learning the new theory, and we are strongly convinced of the need to investigate it empirically. Unfortunately, research on the relation between student understanding of an older theory and difficulties in learning its successor is currently scant.

**2.1 Cognitive Factors**

The major difference we notice between the knowledge of a framework theory and that of a scientific theory is that the latter involves the ability to understand and use different public representations of relevant concepts – e.g., linguistic, mathematical, diagrammatical - for descriptive, explanatory, and problem-solving purposes (Arabatzis, 2020). It follows that shifts in ontology, causality, practices due to theory change may affect all these public representational modes, depending on the relationship between the two theories. A mode of representing a scientific concept may be articulated into different sub-modes. For instance, the linguistic mode includes at least a definitional aspect (Margolis & Laurence, 2021) and a metaphorical one (Amin, 2015) that undergo different changes (see Section 3.1).

Another aspect that is always associated with the knowledge of a scientific theory is the understanding of its exemplars, that we intend in a Kuhnian sense as "concrete exemplary problem solutions" (Hoyningen-Huene, 1993, p. 160). They contribute to the definition of the concept of "paradigm" developed by Thomas Kuhn, and may refer in general to what is viewed as a scientific problem within a theory and what corresponds to an acceptable solution according to the scientific community. For our purposes, what matters is that exemplars "include for one, those solutions encountered by students in the course of their training, in lectures, exercises, laboratory assignments, textbooks, and so on" (Hoyningen-Huene, 1993, p. 134). Theory change is always accompanied by a change in exemplars, which may be strongly context dependent, giving rise to fragmentation in the learner's knowledge structure on the new theory.

**2.2 Epistemic Cognition and Metacognition**

Personal epistemology is an intricate topic, as there is a variety of different perspectives on phenomena of epistemic cognition (Sandoval, Green, & Bråten, 2016).

One of the primary ways in which the subject has been described is "epistemology as a system of beliefs that are explicitly multidimensional and function more or less independently of one another" (Hofer & Bendixen, 2012, p. 230). Sophisticated beliefs on scientific knowledge and knowing could be summarized as follows: knowledge is tentative and evolving rather than certain and fixed, complex and interconnected rather than fragmented, constructed by people rather than perceived in nature, justified by appeals to evidence and coherence rather than authority. The roots of this view are domain-independent, although issues of domain specificity are being increasingly examined: e.g., professional mathematicians could argue for epistemological absolutism concerning mathematical truths as a sophisticated view (Sandoval, 2016). Nevertheless, we are not aware of studies investigating theory-specific issues from this perspective, which is therefore out of our scope.

A theoretical framework that is well aligned with a discipline-specific, situated view is the contextualist resource framework of Hammer and Elby, focusing on epistemic cognition in response to situational demands (Elby & Hammer, 2010). Over time, the activation of epistemic resources in a specific context may form a stable network corresponding to a sophisticated belief, i.e., an "epistemological frame". Studies on epistemological framings during the learning of successive theories have been made exactly in the context of the classical-quantum transition. Dini and Hammer (2017) have conducted a case study on a single successful student in QM. They find that, while framing mathematics as expressing physical meaning is needed for success with classical physics, in QM it is essential since common sense cannot be of help. Consequently, they suggest that courses should support students in their sense-making at this level. A large amount of information has been provided by Modir et al. (2019) on the mapping of published difficulties in learning QM into errors in epistemological framing and resource use, identifying different categories of unproductive framing and transitions. However, the authors do not specify whether it is possible to find a link between the learning of QM - or of specific quantum topics - and patterns of errors in framing. Therefore, the question remains open whether the grain size of analysis within this framework is suitable to identify significant regularities at the level of theory change.

A third way to explore epistemological issues in CC from a theory to its successor might consist in focusing on scientific epistemological differences between these theories and studying the alignment between the personal and scientifically acceptable views on the new theory after instruction. For instance, Baily & Finkelstein (2010), have investigated university student perspectives on quantum physics, and find that their opinions often parallel the stances of expert physicists. Not surprisingly, they find a number of different views, since the scientific epistemology of QM is controversial. In addition, in the process of learning QM, students may encounter conceptual shifts which require them to revise very basic tenets about nature, for example renouncing the idea that all microscopic objects are in a well-defined place at a given time. It is possible, and sometimes reported in the literature (Ravaioli and Levrini, 2017; Authors_c) that this requirement has complex interactions with students' personal systems of beliefs, resulting in effects on motivation to learn, and even on affective factors (strong repulsion or high fascination towards the new theory), which may hinder, or in some cases stimulate, the process of learning. It is possible that these effects are related to the need of a revision of fundamental principles of nature at a time, with respect to the development of the individual, when a full network of personal beliefs and values has been built around such principles, or at least includes them. There is ongoing research on these issues, and we will not delve further into them in this article.

Moving on to discuss metacognition, while a framework theory is not subject to metaconceptual awareness, we must assume that knowing a scientific theory implies a certain degree of metaconceptual awareness of disciplinary content and related problem-solving. This awareness may be lower in relation to newly presented content. However, it is not clear whether and how there can be a difference between the two kinds of CC in terms of metacognitive development.

**2.3 Affective, motivational, and sociocultural factors**

Motivation depends on many circumstances, in the first place on the stage of education we refer to (higher education as opposed to school), and the same can be said for sociocultural factors. Both can vary with the educational environment designed and implemented by the relevant instructors and institutions. Analogous considerations are valid also for affective factors. However, with the exception of the aspects discussed in Section 2.2, none of these factors appears to be related to specific theory changes, and therefore they are outside the scope of this work.

**3 Characterizing the General Features of the Classical-Quantum Conceptual Change**

Here, we use the transition between CM and non-relativistic QM as context for a further development of the model outlined in the previous section. Our aim is to characterize in detail the dynamics that may affect the factors involved in CC to a successive theory, with a special focus on the change in concepts at various public representational levels. This will allow us to build the instruments needed for exploring the connections between specific forms of change and learning challenges elicited in literature.

Overall, no researcher or instructor argues that the acquisition of quantum knowledge is meant to displace the classical one. Both theories work just fine in their respective fields of application, and the possibility of using QM in most classical contexts represents an unsolved issue in the foundations of physics (Schlosshauer, 2007).

Therefore, for the development of a quantum understanding, there is a need to reconfigure existing knowledge acquired in the learning of classical physics, but the new knowledge is supposed to be integrated with the old one.

**3.1 Changes in Concepts at Three Representational Levels**

Theory change deeply affects semiotic resources as regards discipline-specific language, mathematical constructs, and visual tools. Consequently, we need to examine how scientific notions change at three representational levels that we denote as "quasi-qualitative", "mathematical", and "visual".

The first level involves the description of basic terms of CM and QM by means of discipline-specific language. We choose to use the expression "quasi-qualitative" because in QM a purely qualitative description of a physical concept may not be possible at all: e.g., in order to discuss the intuitive idea of state of a system within a wave approach, we cannot do without the mathematical structure of probability distributions. At this level, changes include two different aspects of the linguistic representation of scientific concepts: a definitional one (Margolis & Laurence, 2021), concerning constituents of basic terms such as physical quantity, measurement, and state, and a metaphorical one, dealing with conceptual metaphors in physicists' speech and writing (Amin, 2015). Metaphors used in QM employ images and situations taken from everyday experience and CM, but assign them new meaning: e.g., a physicists may speak of a quantum particle "leaking through a barrier" without ascribing to the particle the full ontology of a liquid dripping from a container. The impact of new conceptual metaphors on student difficulties in learning QM has been studied by Brookes and Etkina (2007). Changes in the definitional structure of the concepts can be denominated as "ontological", those in conceptual metaphors, instead, as "metaphorical". Although CC research from naïve to scientific knowledge already includes the investigation of changes at a linguistic level, the case under scrutiny presents a significant difference: prior knowledge is about concepts of a scientific theory that are developed in the context of formal instruction.

Changes at a mathematical level, instead, represent a totally new feature of this type of CC, which is absent in the transition from naïve to scientific knowledge. While naïve physical knowledge is not mathematized, CM employs a sophisticated mathematical language. After years of systematic instruction in physics, students are expected to have acquired this language and its physical interpretation. Tracing CC in the mathematical representation of physical notions requires a conceptualization of the constructs of pure mathematics and of their role in physics. According to Sfard (1991), a mathematical construct can be seen both as a process (specifying its input-output relationship: e.g., from the terms in a vector sum to its resultant) and as an abstract object (focusing only on the output at a global level: e.g., a free vector as a mental embodied object). In this work, the term "object" or "process" will be applied to a mathematical construct depending on its role in a physical theory, and in particular depending on whether the processual application of the construct admits a physical interpretation. Indeed, physics makes extensive use of the mathematical representation, both for computational and structural reasons (Uhden et al., 2012), the latter meaning that mathematical constructs supply structure to physical concepts and situations, and act as mediators in the process of developing an understanding of these concepts. In introductory QM, substantial changes concern constructs already known to students (see Section 4.2): new versions of them are introduced for the representation of basic terms of the theory (e.g., a new type of operator, the Hermitian one, for representing physical observables). The educationally relevant terms of comparison are the classical version of a construct students are familiar with, and its quantum version. Take the notion of vector intended as a mathematical object having magnitude and direction: in formal instruction on CM, this construct is primarily used to represent physical quantities and lies in the Euclidean space. In QM, instead, vectors typically represent states of physical systems, are abstract unit vectors, defined up to a phase, and lie in a complex Hilbert space. In addition, the classical and quantum features of the processual counterpart of vectors as mathematical objects, i.e., vector superposition, are radically different from one another (Authors_b, and Section 4.2). Of course, abstract complex vectors already existed in classical physics and may be known to students (e.g., phasors), but since they represent a marginal topic in introductory curricula, it is reasonable to assume that their existence affects neither the idea nor the physical meaning of a vector developed by students in the learning of classical physics.

The transition from CM to QM involves (at least) four types of change at the mathematical level. We illustrate this fact with examples taken from the case at hand. First, changes at a global level: in the conceptualization of a construct (object, process, or both – see Section 4.2 for shifts in the conceptualization of operators) and in its referent (e.g., quantum vectors represent system states, not physical quantities). For mathematical processes, the referent includes their purpose and procedure. Second, change in the role of its constituents (e.g., unlike classical vectors, a positive or negative direction of a state vector has only a conventional value similar to the choice of a reference system). Third, change in notable instances of the construct (e.g., the new notions of eigenvector of a Hermitian operator and of tensor-product vector) and notable physical situations described by it (e.g., systems with a definite value of an observables, and separable/entangled systems). Last, change in other features that concur in shaping the structural role of the construct (e.g., a state vector is an element of a Hilbert space, not of a Euclidean one). In general, a shift in the ontology of the mathematical construct turns into further representational changes as regards its role in physics: the physical implications of perpendicularity of Euclidean vectors are different from those of orthogonality of Hilbert space vectors. This change gives rise to deep learning difficulties, as we will see in Section 4.2. New constructs that have no equivalent in the older theory can also be characterized in terms of the same four factors. Since physics (and science, at large) is not interested in mathematical constructs as purely mathematical objects and processes, but in their role in mediating the representation of a scientific concept, we interpret these changes as "representational".

Moving on to examine the visual level, in QM we can visualize experimental setups in the lab, their simulation in educational software environments (e.g., OSP-Spins[2] and QuVis[3]), and mathematical constructs. As regards non-relativistic QM, the most popular representation in traditional teaching is the diagram of the real part of stationary wavefunctions in the context of step potentials. In order to analyze changes in this visual construct, we can use the same description of change developed for the mathematical level, adapting it to a different representational mode: change in global features (conceptualization and referent), change in constituents of the construct, change in notable instances of the construct and notable physical situations described by it, and change in other features that take part in shaping the structural role of the visual construct. In science, visual constructs that represent mathematical ones establish a further level of mediation: the representation of a scientific concept is mediated by a mathematical construct, whose description is in turn mediated by a diagram. With respect to the role of visual constructs in science, the changes are clearly of representational nature.

**3.2 Changes in Exemplars**

Tracing educationally significant change in exemplars is a multifaceted and difficult task, given the wide variety of contexts in CM, of textbook exercises and related resolution strategies, as well as laboratory assignments, and so on. As regards the teaching of QM at secondary school level, while there seem to be an agreement among experts on which topics are considered to be important (Krijtenburg-Lewerissa et al., 2018), the debate remains open on the inclusion of exercises and laboratory assignments (Krijtenburg-Lewerissa et al., 2017). At upper undergraduate level, instead, a comparison of classical and quantum exemplars can be based on the nature of textbook exercises and their resolution strategies. An educational analysis (Authors_e) shows that both QM textbooks and educational research mainly focus on the following tasks and related subtasks: finding 1) the results of the measurement of an observable on a quantum state, 2) the time evolution of the state, and 3) the time evolution of the probability distribution of an observable on a state. The first and the third task have no equivalent in CM, since in a classical context measurement is more about experimental techniques and issues rather than ideal processes in theoretical physics, and physical quantities are assumed to have definite values on a system. The second might call to mind state transformations in thermodynamics, and the time evolution of the state in Hamiltonian mechanics (upper undergraduate students). In order to assess whether it represents a familiar task, we need to discuss how we get to the result.

---

[2] Belloni et al. (2007).
[3] Kohnle et al. (2015).

According to Authors_e, the net of notions involved in the resolution of all three tasks is the following: the closely connected concepts of dynamical variable (observable in QM) and relevant relations (compatibility and incompatibility), measurement, state and eigenstate, time evolution in the absence of measurement. The result is found by analyzing mathematical expressions attached to these concepts: the superposition of a given set of eigenvectors, or commutation relations (evaluating compatibility and incompatibility of a given couple of observables). Subordinate procedures often needed to obtain these expressions are the change of basis, the identification of energy eigenstates and eigenvalues for the system under scrutiny, the application of the time evolution operator.

As we have seen in Section 3.1, and will discuss in detail in Section 4, all of the involved concepts have undergone an ontological change in the transition from CM to QM. Some mathematical objects and processes a representational one. Others, such as the commutator, is an operator that results from a composition of two operators, and is used in the first place to assess the compatibility and incompatibility of observables, but also in algebraic procedures needed to identify their eigenstates and eigenvalues. In addition, the connections between concepts have changed (e.g., QM establishes an unbreakable bond between the concepts of state and ideal measurement).

In summary, the novelty of two fundamental tasks and of the resolution strategies of all three tasks might account for the lack of a robust knowledge structure and the strong context dependence of productive reasoning reported by research on upper undergraduate students after intensive instruction on QM (Marshman & Singh, 2015). Many students come to see QM just as a collection of mathematical rules (Johnston et al., 1998).

**3.3 Issues in the Scientific Epistemology of Quantum Mechanics**

Since scientific epistemology represent a controversial topic in QM, that may affect what we consider shifts of ontological or epistemological nature, in this article we limit ourselves to discuss the relation between scientifically acceptable perspectives and instruction. As a matter of fact, the nature of quantum systems described by the mathematical formalism, the extent of the information we can get on them, and the explanation of the observations in the lab depend on the chosen interpretive stance. There exist several contending schools of thought over this subject and a corresponding number of interpretations of the theory. See, e.g., Home and Whitaker for a review of statistical interpretations, Dieks and Vermaas (1998) for modal interpretations, Bub (1997) for no-collapse interpretations and more. Some textbooks based on these alternative views have been published in recent and in less recent years (e.g., Ballentine, 2014; Griffiths, 2001) but they are indeed more suitable to graduate courses rather than introductory ones. Nevertheless, to date no alternative view has prevailed over the mainstream approach (e.g., von Neumann, 1955). This interpretation can be briefly outlined by means of three statements:

1. a pure state provides complete information on the behavior of an individual quantum system;
2. an observable has a determinate value if and only if the quantum state is an eigenstate of the observable;
3. the quantum description of processes includes two different types of state evolution: in the absence of measurement, the unitary evolution governed by the Schrödinger equation; in measurement, the evolution prescribed by the projection postulate.

The last statement is the most controversial one and can be considered unsatisfactory for a variety of reasons (Bub, 1997). However, in a survey on 27 faculties from different US universities, 24 agree that the projection postulate is an essential part of QM and needs to be taught and emphasized in the course, even if seven of them suspect it is 'incomplete' or 'wrong' (Dubson et al., 2009). This consensus is accounted for in educational studies on QM, where great importance is generally ascribed to student ideas on the effect of measurement on the quantum state (e.g., Goldhaber et al., 2009; Zhu & Singh, 2012; Gire & Manogue, 2011; Passante et al., 2015a). If we accept the standard interpretation, epistemic issues turn into ontic ones. The question is no longer: "what can we know about the quantum world?", but rather "how is this world really?". While the completeness of information on systems provided by QM is rarely discussed in educational studies and curricular materials, in all other respects most of them operate within the Schrödinger picture of the standard interpretation of QM. Consequently, in the rest of this work we will adopt this perspective.

**4 Exploration of the Connections between Theory Change and Challenges in the Learning of Quantum Mechanics**

In this section, we examine the results of existing research on challenges in the learning of QM through the lens of the types of conceptual dynamics described in Section 3.1. The goal of the analysis is to identify solid connections between changes in concepts at quasi-qualitative, mathematical and visual level and cognitive demands[4] of learning QM. In this process, we aim to develop a comprehensive picture of the educational significance of this aspect of theory change (Purpose 1) and to explore the validity of the analogy between challenges in learning QM and introductory physics (Purpose 2).

By reading the relevant literature, it is immediate to see that student difficulties generated by ontological and representational changes are intertwined with those generated by change in exemplars. As a matter of fact, the existence and the format of internal representations in students' mind is inferred by external indicators in the performance of tasks that often include theory-specific exemplars. In addition, the solution of exercises has clearly a part in forming students' ideas on the concepts at hand. Although we are aware of the impossibility to fully disentangle the cognitive effects of these two types of change, here we choose to analyze empirical results primarily from an ontological and representational perspective. A comprehensive analysis of the educational impact of changes in exemplars and epistemic changes will be the subject of future woks.

As specified in Section 3.1, the terms of comparison between the two theories are not chosen according to the historical reasons for the change under scrutiny, nor according to an abstract comparison between the conceptual structures of the two theories, but based on analogous entities students are supposed to be familiar with after a course on CM and are likely to encounter in a standard course of QM. However, it is not a trivial task to decide what we consider as a relevant "concept" and what a property of this concept. For instance, a physical quantity and a vector are concepts, but also incompatibility (of physical quantities) and modulus (of a vector) are. As regards entities taking part to the classical and/or the quantum physical description, we choose to denominate as "concepts" the basic conceptual instruments used for the description of a physical system (physical quantity, measurement, state, time evolution, general model), as "constructs" their formal representations and fundamental mathematical processes used to get information from or on the world (vector, vector superposition, wavefunction, operator), and the visual representation of systems and mathematical constructs (system diagram, wave diagram).

A detailed description of *ontological* dynamics in the concepts and *representational* dynamics in the constructs under scrutiny is reported in tables presented in the supplementary file titled "Dynamics of Theory Change from CM to QM". Trajectories of individual concepts and constructs are classified as follows: new formation, evolution, disappearance (Arabatzis, 2020). The same is valid for the trajectory of each of their properties. Properties of constructs are subdivided according to the taxonomy presented in Section 3.1 (global features, basic constituents, notable instances and notable physical situations described by the construct, other features shaping its structural role).

---

[4] We denominate as cognitive demands all those *general cognitive tasks required for learning a specific concept*.

The protocol of the analysis is the following: for each change in one of the selected notions (see tables), we scan empirical studies in the search for common difficulties in the interpretation and use of the new element (if it has formed in theory change), of its quantum version (if it has evolved), or to recognize its absence (if it has disappeared). Each representational level is discussed in a separate section. Within each section, all issues concerning changes in a given notion are presented in a separate paragraph. We are aware that concept building is often a result of the interplay of different notions and, when students are exposed to multiple representations, of the interplay of these representations. Therefore, we point out when learning issues are clearly related to the first and/or the second interaction. In the first case, we report the issue in the paragraph concerning the last of the notions involved. For *metaphorical* changes, we refer to Brookes & Etkina (2007). Since the reported impact of metaphorical change in the learning of QM concerns the interpretation of visual constructs, it will be mentioned in the corresponding section. In Section 4.4, we summarize the general nature of conceptual dynamics and assess, from a global perspective, its impact on challenges already known to the scientific community.

**4.1 Quasi-Qualitative Level**

At this level, we consider only changes in the definitional aspect of the quasi-qualitative representation of the concepts of *physical quantity*, *measurement*, *state*, *time evolution* (in the absence of measurement) and *general model*.

A discussion of research results at this level requires special care, since student concept-building on the basic terms of the theory may be influenced by their mathematical description. Although less numerous, data collected on lower undergraduate students and on secondary school students are extremely valuable, given that the first population has not been exposed to the full mathematical machinery of the theory, and the second one often not even to its basic constructs. As regards results on upper undergraduate students, we only include naïve or hybrid ideas that are clearly unrelated to the mathematical world of QM.

Difficulties on *physical quantities* are evidently a result of the trajectory of this concept. Secondary students tend to ascribe classical-like features to quantum observables, and even struggle to accept the possibility of quantities having no definite value. The issue can be illustrated by means of the following examples reported by Ravaioli and Levrini (2017): during individual interviews, students stated that "The object itself does own a well-defined property, that's what I believe. […] As Einstein's, mine postulate is that an object has to embody well-defined properties", or "I believe objects to have a definite position and momentum. There is something that escapes our understanding" (see also Authors_c). While all unconstrained classical quantities have continuous values, some observables present only a countable (e.g., angular momentum) or finite (e.g., spin) set of values. Documented difficulties in interpreting this transition involve also the concept of state and are discussed in the paragraph devoted to this concept. Incompatibility of observables is one of the most remarkable new features of QM. While this notion and its implications are challenging for students (see, e.g., Singh & Marshman, 2015), questions concerning incompatibility generally involve observables, measurements, states, and possibly also time evolution at the same time, and include mathematical notions, such as vectors, wavefunctions, etc. Issues with incompatibility will be presented in Section 4.2.

Not surprisingly, interpreting the stochasticity of quantum *measurement* and, as a result, quantum uncertainty, represents a strong challenge to students. Ayene et al. (2011) identified four categories to describe student depictions of the Heisenberg uncertainty principle: 1) uncertainty is mistakenly interpreted as a measurement error due to an external effect, 2) uncertainty is wrongly described as a measurement error due to the limitations of the instrument, 3) uncertainty is falsely thought to be caused by measurement disturbance, and 4) uncertainty is correctly seen as an intrinsic property of quantum systems. Only 3 of 25 second year students fell in the fourth category. Independently of the approach of the course, similar inconsistent interpretations are reported in other investigations: wave approach (Müller & Wiesner, 2002) and Feynman's sum over paths (Authors_a).

As we argued in Section 3.1, the concept of quantum *state* has an inbuilt mathematical nature. This notion and those relying on it, e.g., quantum superposition, collapse, and time dependence, are classified by Krijtenburg-Lewerissa et al. (2017) as "complex quantum behavior". Few data are available on secondary school students and lower-undergraduate ones. At a quasi-qualitative level, difficulties with the concept of state are reported in the context of polarization: asked whether the quantum state is a physical quantity or not, 6/32 secondary school students unproductively activated the conceptual resource that the state concerns measurement (the state "is measurable"), thus identifying it as a physical quantity (Authors_b). Within this context, difficulties emerge even at upper undergraduate level. During interviews reported by Singh and Marshman (2015), students argued that "since a polarizer can have any orientation and the orientation of the polarizer determines the polarization state of a photon […], it did not make sense to think about the polarization states of a photon as a two-state system" (p. 5). These difficulties invest the evolution of the concepts of state and physical quantity: both cases highlight issues with the binding link between the quantum state and the measurement of quantities – a link that was absent in CM; the second one, in addition, shows that at least in some contexts and with some educational approaches, interpreting the transition from continuous (classical) to discrete (quantum) sets of values may be a challenge to students.

A treatment of the *time evolution* of quantum systems is usually outside the scope of secondary school curricula. However, some of its elementary features may be treated at a purely qualitative level: the disappearance of the point-particle form of time evolution (trajectory) and the introduction of a wave propagation. Difficulties of secondary school students with these implications of theory change also concern the general model of quantum system and are reported in the next paragraph. At the upper undergraduate level, Passante et al. (2015a) identified two naïve ideas on time evolution after a measurement: revival (the state returns to the form it had before the perturbation) and decay (from an excited state to the ground state). These ideas appear to be unrelated to the mathematical structure of the theory. Since the concept of excited state has no place in non-relativistic QM, the second difficulty might be derived from the exposure to popular science content. Evidently, understanding the features of the quantum time evolution is a considerable challenge to students at all levels.

Last, it is well known that the description of the *general model* of a quantum system is rather puzzling to students. Krijtenburg-Lewerissa et al. (2017) devote a specific section of their review to the illustration of difficulties with the wave-particle duality. One example: several secondary school students and lower undergraduate students interpret the wave behavior of electrons as that of a pilot wave, which forces the electron into a sinusoidal trajectory. Such beliefs represent a hurdle also in the understanding of stochastic interference: Kreijtenburg-Leverissa et al. (2017) report that, in interpreting the results of a double-slit experiment, some secondary school students considered photons to deflect at the slit edges and move in lines towards the screen.

**4.2 Mathematical Level**

The mathematical constructs we consider are the following: *vector, vector superposition, wavefunction, operator*. Changes may involve their global features, the representational or conventional role of its constituents, notable instances of the construct and physical situations described by it, and other features that have implications on the structural role of the construct.

Empirical studies on student understanding that adopt a mathematical representation – mostly at upper undergraduate level or above – suggest that learning issues are not a result of the technical sophistication of the machinery of the theory, but of the difficulty to activate mathematical sensemaking in QM (Wan et al., 2019). Already in early investigations, researchers noticed that, when learning QM, students attempted to equate clearly different algebraic expressions for physical reasons, an issue which would not have been raised in a mathematics course (Singh, 2007). Later studies proved that many issues with math concern its structural role. Wan et al. (2019) showed, e.g., that students "are proficient with distinguishing the modulus squared of complex numbers, and yet they often do not apply this skill correctly when comparing quantum probabilities. As a result, students often fail to recognize the measurable effects of different relative phases" (p. 11).

Research on learning challenges with *vectors* in QM evidenced difficulties with phases (a new constituent), the notable notion of eigenvector (absent in introductory physics), and their new space of definition (a feature that affects the structural role of the construct). As regards phases, the subtle distinction between the conventional role of the overall phase and the representational role of relative phases in superposition states generates several issues. They are discussed in the next paragraph on vector superposition. A consistent understanding of the notion of eigenvector is crucial for acquiring the basics of non-relativistic QM, but research showed it is very difficult to achieve. There is plenty of studies on the topic, but most of them involve at the same time the constructs of vector, superposition, and operator. For this reason, we report their results when discussing difficulties with the last construct (operator). The confusion between classical and quantum vector spaces represents a deep issue. Classical vectors lie in the Euclidean space (lab space), whence the idea that a vector with a label $x$ is orthogonal to a vector with a label y, or cannot influence it, as often happens to components of classical vector quantities along the axes (Singh, 2008, Singh & Marshman, 2015). In the context of spin-half particles, students ascribe these properties to quantum state vectors, erroneously claiming that spin-x vectors are orthogonal to spin-y vectors and that a measurement of spin-x has no influence on the state of the system as regards its spin on the y axis.

The trajectory of *vector superposition* is related to the onset of various types of difficulties, e.g., with the change in referent, the role of new constituents (overall and relative phases), and the new notable instance of entangled superposition. For what concerns the referent, while classical superposition of forces and waves often involves the composition of more physical entities to find the resultant, quantum superposition is only a decomposition of one physical entity to obtain info on measurement. Passante et al. (2015b) investigated student ability to distinguish superposition states (decomposition of one entity) from a mixture of states (composition of more entities), that can be classically interpreted as lack of knowledge about the system. They found that even students at the beginning of graduate QM instruction struggle in a similar way as undergraduates: 55% of 31 students did not discriminate between the two types of state. Student beliefs on relative phases have been studied by Authors_d and Wan et al. (2019). The first reported that in a question in the context of spin, just six out of 32 3rd year students recognized that, without specifying the relative phase in a superposition, the knowledge of the state is incomplete. Of those who did, only half identified the measurable effects of relative phases in a follow-up task. Based on these results and on interviews of a subset of students, the authors observed that most of them neglect phase difference, and even part of those who do not, either interpret it as a needed formal element, but without physical meaning, or as the overall phase. The second study asked students isomorphous questions in different contexts or disciplines (mathematics vs. physics), and concluded that, while answering correctly to the mathematical question, half of 86 junior level students did not recognize that particles with different relative phases are experimentally distinguishable. Student understanding of superpositions of tensor product states has been investigated by Kohnle and Deffenbach (2015), finding various patterns of difficulties: e.g., some students incorrectly claim that particles in a product state must have definite values of the entangled observables, others that a product state is always an entangled state. Studies on consecutive measurements on a superposition, involving eigenvectors, the distinction between different sets of them, and incompatibility, are discussed under operators.

With the term *wavefunction,* we refer to the mathematical representation of a wave, classical or quantum. A number of challenges with the quantum notion of *wavefunction* and with its use to describe quantum states (e.g., the new role of constituents such as the amplitude) have been elicited by having students sketch and analyze wavefunction graphs, and therefore will be discussed in Section 4.3, devoted to learning issues at a visual level. Difficulties emerging at a purely mathematical level concern the fourth type of change: new constraints for the admissibility of a wavefunction and its decomposition into energy eigenfunctions. In a written survey, Singh (2008) asked 202 graduate students to assess whether examples of wavefunctions provided by the researcher represented allowed states for an electron in an infinite energy well and interviewed a subset of students. Only one third gave a correct answer in all three cases. While students were able to use boundary conditions in order to assess whether a wavefunction is allowed, many of them added excessively restrictive conditions for the admissibility of a wavefunction: either that it must be an energy eigenfunction, or that it must be possible to write it as a linear superposition of a finite number of these functions. Actually, all single-valued $\mathbb{C}^1$ square summable functions defined in the domain are admissible and, however, a linear combination of wavefunctions may be compressed into an equivalent expression where the original basis states are not visible at a first glance. As a result, students may not recognize it as such. This happened to half of the students.

One of the most important but less understood constructs in QM is the Hermitian *operator* and its relationship with all the other concepts and constructs discussed in this analysis. Difficulties elicited by research concern global features of the construct (the conceptualization of Hermitian operators as an object, since its action has no physical meaning), a new notable instance of operator (the Hamiltonian one, representing energy), and the new notion of eigenvector of a Hermitian operator, whose understanding affects predictions on the results of measurement and on the time evolution of an eigenstate of an observable (that requires to evaluate whether this eigenvector is a simultaneous eigenvector of the Hamiltonian operator). The first difficulty is connected with the need to move from a conceptualization of an operator as a mathematical process, that is familiar to students, to that of an operator as an object: although the action of most Hermitian operators on a vector has no physical meaning, many students develop the idea that when an operator representing a physical observable acts on a state vector representing a system, it corresponds to a measurement (e.g., Singh, 2008; Goldhaber et al., 2009; Gire & Manogue, 2011, Singh & Marshman, 2015). This belief is so strong that students neglect the often-repeated fact that measurement is an intrinsically stochastic process, and therefore no equation can predict its result. A well-known difficulty with the role of the energy operator in governing system dynamics, and in particular with its unique property of changing eigenvalues and eigenstates from context to context, is the interpretation of the eigenvalue equation for the Hamiltonian: many students interpret this relation not as the definition of a mathematical task, but as a fundamental law of physics. In general, the interplay between the vector structure of states and the operator structure of observables established by QM represents a major challenge to students. This is proven by many studies on the interpretation of the fundamental concept of eigenvector which - as noted by Authors_e - represents the junction between the two structures, as it is a state vector, but at the same time a physical piece of information encoded in the Hermitian operator that represents an observable. Common difficulties concern the distinction between eigenvectors of different Hermitian operators (Singh & Marshman, 2015), and include, e.g., the belief that eigenvectors of operators corresponding to all physical observables are the same. These issues have dramatic effects both in predicting the possible results of measurement and in time evolution. See, e.g., the belief that a position measurement on an energy eigenstate does not change the energy of the state and therefore that a subsequent energy measurement will give this value with certainty (Passante et al., 2015a), and the belief that when the system is in an eigenstate of an arbitrary observable, it will stay in that state until measured (Goldhaber et al., 2009; Singh and Marshman, 2015). Another consequence of the inability to discriminate between sets of eigenstates of incompatible observables, and possibly also of the confusion between Hilbert space vectors and Euclidean ones is misinterpreting or ignoring the change of basis formulas (Authors_d).

**4.3 Visual Level**

Any consideration on visual representations must start from the fact that "quantum systems do not admit any visualization […] by means of familiar images such as an image representing the atom's planetary model. At best, they can be described by graphical representations showing mathematical properties […] such as Feynman diagrams" (Levrini & Fantini, 2013, p. 1898). A major change from CM to QM corresponds to the disappearance of *system diagrams,* such as the free body diagram, and in general any visualization of systems and their evolution. This change deprives students of important resources in organizing scientific knowledge and entails a much higher degree of abstraction as compared to introductory physics (Marshman & Singh, 2015). The impossibility to visualize quantum systems obviously represents a hurdle in building a mental model of their features, as attested by students using trajectories to describe the motion of a particle (see Section 4.1), and enhances the importance of mathematical sense-making.

Challenges with *wave diagrams* elicited by literature concern basic constituents (the new interpretation of the vertical axis), notable instances (new nature of bound and scattering systems) and physical situations (a new phenomenon: tunneling), and other features (new boundary conditions). Research on student difficulties with *wave diagrams* has been particularly active from 2000 to 2010. Students were asked to sketch, given the potential energy diagram and particle energy, the graph of the corresponding stationary wavefunction, or to interpret graphs provided by the researchers. Open questions and interviews were used to deepen the insight on student conceptions. The most notable difficulty was elicited primarily in the context of tunneling: various researchers found that a large number of students expressed the classical-like idea that the amplitude of the wavefunction is related to energy, not probability. Since the amplitude of the transmitted wave is lower than that of the incident one, they concluded that the particle loses energy in tunneling (e.g., Wittman et al, 2005; McKagan et al., 2008; Robertson & Kohnle, 2010). A typical student statement is the following: "tunnel exponentially lowers the electron's energy so when it emerges it has lower amplitude of oscillation" (Robertson & Kohnle, 2010, p. 268). Etkina and Brookes (2007) claim that a major contribution to the onset of this idea is provided by student interpretations of experts' conceptual metaphors that ascribe a two-dimensional gravitational ontology to the physical situation. As regards the quantum version of the graphical representation of bound and scattering states in QM, Singh (2008) shows that even graduate students may have difficulties to discriminate between the two types of states. For instance, 8% of 202 students that were asked to sketch the graph of the ground state of a potential well drew an oscillating function in all regions. In the comments, several students displayed confusion about the new meaning of bound and scattering state, and "whether the entire wavefunction is associated with the particle at a given time or the parts of the wavefunction outside and inside the well are associated with the particle at different times" (Singh, 2008, p. 285).

**4.4 Accomplishing Purpose 1 and 2**

The application of this model of CC to the transition from CM to QM has allowed us to build fine-tuned instruments for exploring conceptual dynamics in theory change at different representational levels: quasi-qualitative, mathematical and visual. The basic trajectories followed by scientific concepts result to be the evolution of known notions used to describe systems (physical quantity, measurement, state, time evolution, general model), of their formal representations and mathematical processes used to get information from or on the world (vector, vector superposition, wavefunction, operator), of the visualization of mathematical constructs (wave diagram), and the disappearance of the visual representation of systems (system diagram). At the quasi-qualitative level, we see the new formation of conceptual properties (e.g., incompatibility) and the disappearance of an old property of time evolution (trajectory). At the mathematical level, the constructs at hand undergo changes in global features, in the representational and conventional role of their constituents, in notable instances of the construct and physical situations described by it, and in other features that concur in shaping their structural role. At the visual level, we see the disappearance of system diagrams and the change in purpose, procedure, constituents, and instances of wave diagrams (see related table in the supplementary file titled "Dynamics of Theory Change from CM to QM").

By analyzing existing data on learning challenges in the search for a connection between these trajectories and student difficulties, we are able to draw a detailed picture of cognitive demands related to the aforementioned dynamics (Purpose 1). Concerning the quasi-qualitative level, the cognitive demands that pertain to its definitional aspect are the re-interpretation and use of known concepts in new ways due to theory change, and in particular, the interpretation and use of new unintuitive properties of these concepts (e.g., stochastic interference). At the mathematical level, the change in the global features of known mathematical constructs has been detected as an issue in relation to superposition (only decomposition, not composition) and Hermitian operators (to be conceptualized primarily as objects, not processes). The three other kinds of change have proven to pose serious challenges in learning: cognitive demands involve, e.g., the interpretation of the role of relative phases (constituents of superposition), the interpretation of the new notable notion of eigenvector of a Hermitian operator, the distinction between the structural role of the space of definition of quantum vectors and of classical ones. At the visual level, the intertwined nature of the three representational levels in concept-building is particularly evident. The main cognitive demands on the wave diagram are the interpretation of the role of the vertical axis in the context of a new physical situation (tunneling) and of the representation of bound and scattering waves, that influence ideas on its referents: mathematical (wavefunction) and quasi-qualitative (state). In turn, new conceptual metaphors in the language of instructors and textbooks negatively affect the interpretation of wave diagrams.

Based on this map of the links between cognitive demands and conceptual dynamics in theory change, we are able to explore the boundaries of the analogy posited by Marshman and Singh (2015) between challenges faced by students in learning introductory physics and QM (Purpose 2). According to Marshman and Singh, both student populations display general patterns of difficulty: poor categorization of physics problems, not using problem solving as a learning opportunity, inconsistent and (or) context-dependent reasoning, inappropriate or negative transfer, lack of transfer, "gut-feeling" responses inconsistent with the laws of physics, difficulties in solving multipart problems, difficulties related to epistemological views. However, while at introductory level the cognitive challenges are inherent to the process of revision and integration of naïve knowledge (Vosniadou & Mason, 2012), which is neither mathematized nor socially shared, we find that a significant part of common difficulties in the learning of QM is strongly linked to totally new types of challenges, which result from various forms of change in public representations, including a) mathematical constructs that are already familiar to students, and b) visual representations of these constructs. As we show in Section 4.5 and 5, conceptual dynamics involved in the two types of CC are very different from each other. Therefore, the transition to the knowledge of a successive theory may be favored by the development of qualitatively different strategies (Section 6) that stem from a closer representation of the trajectory of concepts/constructs in theory change (Section 5).

**4.5 Contrasting Conceptual Dynamics in the Two Types of CC**

According to Carey (as cited in Vosniadou, 2008, p. 3), the process of building scientific knowledge requires the re-assignment of a concept to a different ontological category or the creation of new ontological categories (earth: from the category of physical objects to that of astronomical objects[5]). CC from naïve to scientific understanding can also involve differentiation (of *weight* from *density*) or coalescence (uniting of *animal* and *plant* into the new concept *living thing*), as specified in Carey (1999). Chi (2008) focuses on two general kinds of ontological shifts in the learning of physics and biology that she identifies as radical forms of CC: from entities to processes (force), and from sequential to emergent processes (heat transfer, natural selection).

---

[5] This trajectory is shown in detail in Table 2.

In the educational transition from CM to QM, instead, we find that learning challenges are related, in the first place, to the evolution of classical concepts and constructs driven by the appearance of new unintuitive properties (e.g., incompatibility of observables, stochastic interference). Educationally significant trajectories rarely involve a mere differentiation or coalescence of known notions. Even the general model of a quantum system, that at first glance might be described as a coalescence of aspects of a wave model and a particle model, requires a consistent understanding of stochastic interference in order to be properly interpreted. Such shifts in the ontological structure of concepts and in the representational structure of constructs are different from the specific forms described in the traditionally considered cases of CC, but in order to illustrate the nature of these shifts, we need to develop appropriate tools. This is the subject of the next section.

**5 Visualizing Ontological and Representational Evolution by Means of Dynamic Frames**

The analysis presented in the previous section has shown a rich landscape of changes and corresponding challenges in the learning of QM. However, in order to disclose the educational potential of the analysis, we need to design a handy representation that captures its most significant elements, allowing us to devise trajectory-dependent strategies that facilitate the revision of classical concepts/constructs for promoting the understanding of their quantum versions, and to suggest an answer to student need of comparability between CM and QM. For this purpose, we choose to focus the attention on categorical change. Categorization as a learning mechanism has been extensively studied by the educational research community, that has evidenced both its importance and issues related to its use. Vosniadou & Skopeliti (2014) state that "Categorization is the most fundamental learning mechanism, a mechanism which most of the time promotes learning but which, in cases where CC is required, can inhibit it" (p. 4). In the same work, they explain: "Once categorized […] an entity inherits all the […] properties of the entities that belong to the domain" (p. 2). Chi (2008) adds: "categorizing, or assigning a concept to a correct category, is powerful because a learner can use knowledge of the category to make many inferences and attributions about a novel concept/phenomenon" (p. 62). Nonetheless, we are aware that categorical demands are only a subset of cognitive demands: other demands concern the scientific interpretation of conceptual metaphors (Amin, 2015) and the ability to extract data of a situation and process them in order to determine concept-specific information (e.g., the resultant force) across different contexts (diSessa & Sherin 1998; diSessa et al., 2016).

Researchers on CC in learning have used various tools to visualize categorical structures and change. Chi (2008) presents an ontological tree to visualize fundamental categories: entities, processes, mental states, and the internal structure of each one (figure p. 64). Thagard (1992) includes in CC changes in learning as well in the history of science, and uses ontological trees and concept maps to compare the structures of successive theories, e.g., Stahl's phlogiston theory (figure p. 31) and Lavoisier's conceptual system (e.g., figure p. 47). In a table, Vosniadou (2008) presents a comparison of properties of the naïve version of the concept of earth and those of the scientific one (Table 2).

**Table 2** Comparison of students' initial idea of earth and the scientific one (Vosniadou, 2008, p. 5)

**Concept of the Earth**

| *Initial* | **S**cientific |
|---|---|
| *Earth as a physical object* | *Earth as an astronomical object (planet)* |
| Flat | Spherical |
| Stationary | Rotating around its axis<br>Revolving around the sun |
| Supported | Unsupported |
| Up/down gravity | Gravity towards the center of the earth |
| Geometric system | Heliocentric system |

Two tools used for tracing CC in history and philosophy of science deserve special attention because of their clearly formalized syntax: dynamic frames (e.g., Andersen et al., 2006) and conceptual spaces (Gärdenfors, 2000). Dynamic frames provide a visual representation of taxonomic knowledge in terms of a hierarchy of nodes, starting from the superordinate concept and organizing its structure into sets of values (conceptual constituents), each set related to a different attribute (functional concept representing the relation between the superordinate concept and the relevant set of values). Frames are recursive, since each value can be expanded into a new frame that can be specified further by attributes (Gamerschlag et al., 2014), and may display empirical regularities in the form of constraints between attributes and between values. For an example of frame, see Fig. 1. Conceptual spaces, instead, provide a geometric and topological account of concept representation (Zenker, 2014). They are more powerful than frames, since their values incorporate not only concepts, but also ordinal, interval and ratio scales. Moreover, the elements that expand the notion of attribute, denominated as dimensions, can be combined by means of mathematical constraints. In simple situations, spaces naturally collapse into dynamic frames. This tool has been used to model entire theories and theory change (Masterton et al., 2016), but at the cost of losing the visualization of the change in concepts. Since we need to visualize the taxonomic evolution of individual concepts and constructs in a compact form, we choose to adopt and revise the dynamic frame model.

## 5.1 Dynamic Frames

As we said, dynamic frames represent concepts by means of layers of nodes (see Fig. 1). The single node on the left represents the superordinate (label: S) concept (BIRD), all the other nodes represent specific subsidiary concepts. The second layer represents attributes (A) of the concept (e.g., BEAK, FOOT). The third one, values (V) of those attributes (e.g., a BEAK can be ROUND or POINTED). In natural language, each triplet S, A, V represents a proposition: V is the A of S (Gamerschlag et al., 2014).

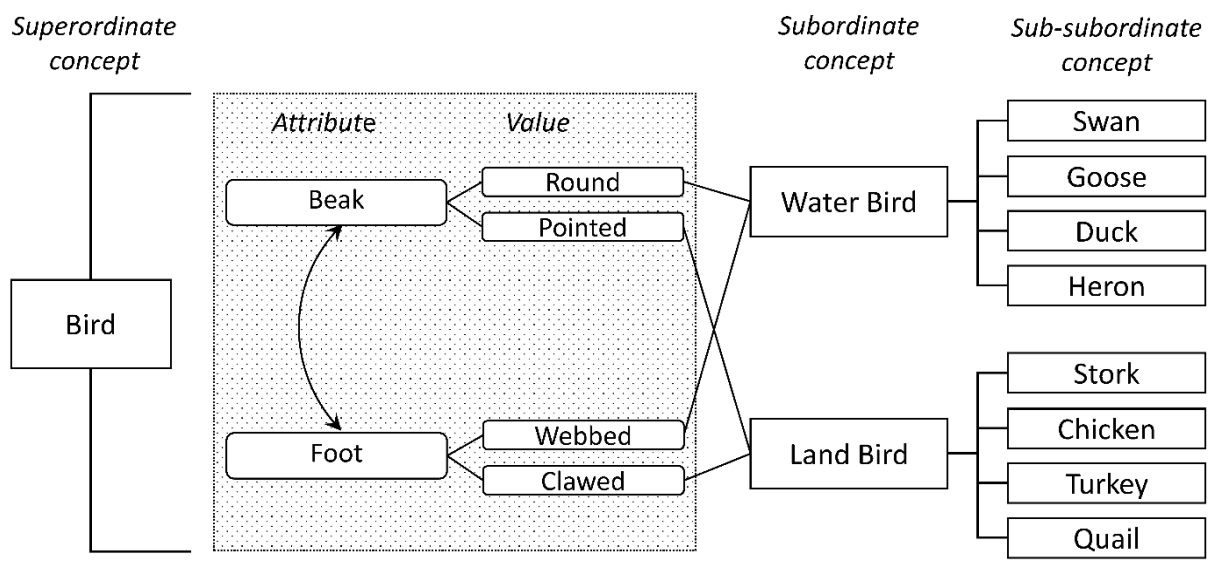

**Fig. 1** Partial frame representation of the Ray taxonomy for birds (Andersen et al., 2006, p. 73)

In the figure, each value is tied to concepts that are subordinate to the same superordinate concept. When a particular subset of values is chosen to represent a specific subordinate concept, values are said to be "activated" (Andersen et al., 2006, p. 44). Similarity and dissimilarity between concepts at the same level is assessed by inspecting the activated values. Two subordinate concepts are similar with respect to an attribute if they have the same value for the attribute, and they are dissimilar if they have different values. As specified in the caption of Fig. 1, frames usually contain incomplete information regarding the superordinate concept, and therefore are denominated as "partial". The choice of the attributes, therefore, is up to the researcher, and is always influenced by "goal, experience and intuitive theories" (Barsalou, 1992, as cited in Zenker, 2014, p. 72).

Andersen et al. (2006) traced CC in the history of science by comparing different frames corresponding to subsequent or competing views of the structure of a scientific concept. See, for instance the classifications for birds represented in Fig. 1 and 2. An issue they extensively address in their book concerns Kuhn's account of CC, with a special focus on revolutionary change and incommensurability. According to the authors, in a frame representation:

> "Incommensurability is a mismatch between the nodes of two frames that represent what appear to be the same superordinate concepts. […] at some point we encounter structures in the two series of recursive frames that do not map onto each other. […] The most serious problems will arise from the addition and deletion of attribute nodes. Incommensurability occurs between two frames for the same superordinate concept when we are confronted with two seemingly incompatible sets of attribute nodes" (Andersen et al, 2006, p. 116).

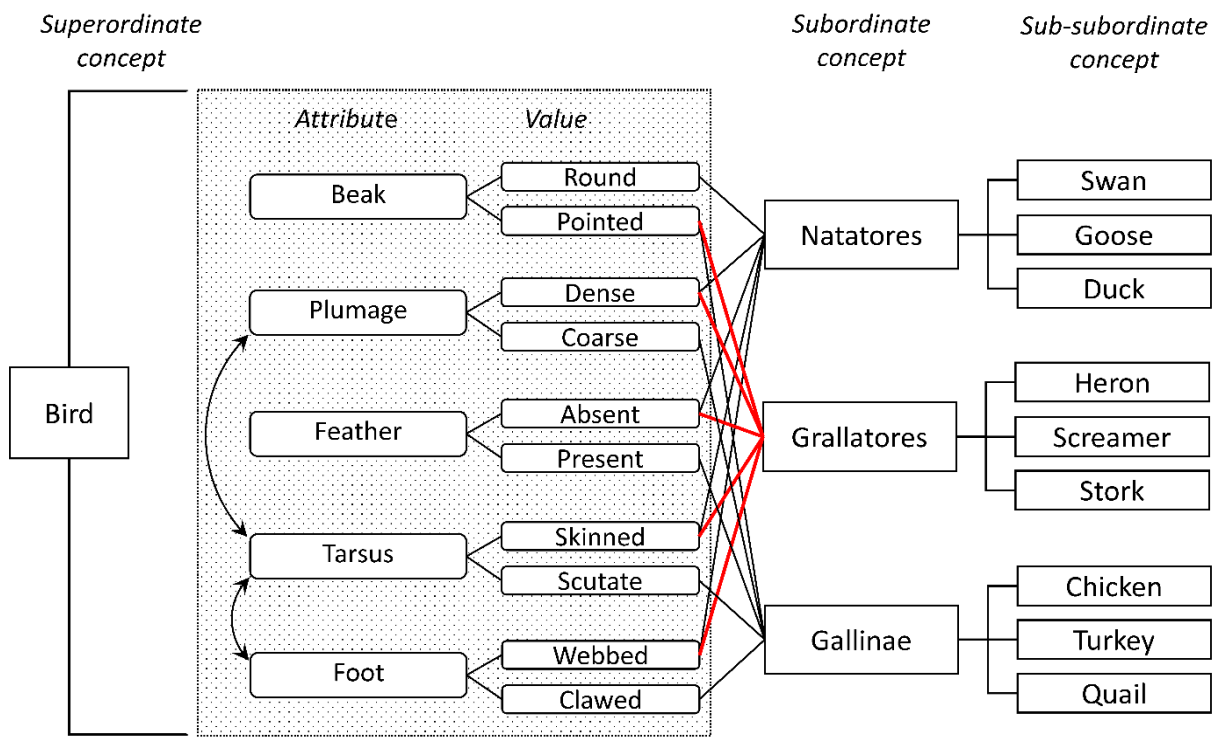

**Fig. 2**   Partial frame representation of the Sundevall taxonomy for birds (Andersen et al., 2006, p. 74). Red lines connect the values to the new category of birds in the frame: GRALLATORES

**5.2 Representing Change and Continuity from Classical to Quantum Mechanics for Educational Purposes**

The general format we use to represent change in scientific concepts and constructs draws on the frame representation used in Schurtz and Votsis (2014). The authors described significant aspects of the transition from the phlogiston theory to the oxygen theory of combustion by using a single frame to express the structure of both theories as regards the concepts of DEPHLOGISTIFICATION / OXIDATION (p. 103) and PHLOGISTIFICATION / REDUCTION (p. 104). Both dichotomies are represented as subordinate instances of the superordinate concept CHEMICAL REACTION, and are differentiated by the patterns of activation. Analogously, we also adopt a color code for value boxes, in order to highlight which one belongs to a theory, to its successor, and to both.

The iconic structure of our frames is displayed in Fig. 3. The superordinate concept is either a basic term of both theories or a construct evolving in the transition from CM to QM. Value boxes are white if they pertain to the classical version of the superordinate concept/construct, black if they pertain to the quantum one, gray to both theories. In the abstract example, the pattern of activation for the classical version includes values 1, 2, 4, 6, 7. For the quantum one, values 1, 3, 4, 5, 7, 8. In some frames we present in this article, it is possible to infer, by means of additional information, the patterns related to specific instances of the notion at hand, e.g., for the SYSTEM QUANTITY frame, the frame of a specific quantity such as position or momentum.

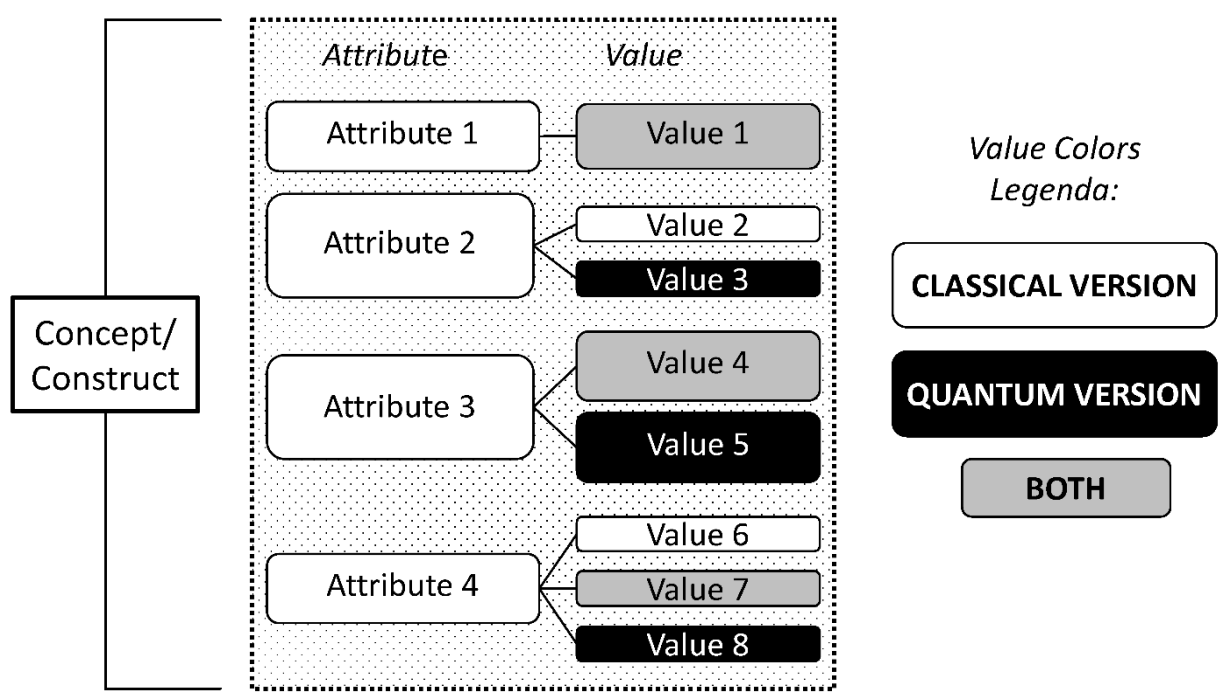

**Fig. 3** The iconic structure of our frames: abstract example

Differently from Schurtz and Votsis (2014), our purpose is not to provide a structural account of theory change, but to represent the structure of the educational transition between theories, and this is reflected on the choice of attributes. First, we only include those attributes that highlight basic or educationally significant aspects of the notions under scrutiny. Second, in order to widen the relevance of this work for researchers and instructors, in the present article we focus mostly on aspects of the transition which are common to undergraduate and high school students, thus avoiding to include composite states (research on high school student understanding of entanglement is just at the beginning), complex phase differences, and eigenstates. The language used for structuring attributes and values is chosen accordingly. Finally, frames designed to describe change in concepts address ontological shifts in their definitional structure. Frames of constructs focus on representational shifts.

However, since the attributes of frames concerning physical concepts are structured to highlight continuity and change at an ontological level, the comparative analysis of the structure of the two theories can play an important ancillary role. It can be proven that many aspects of concept evolution that give rise to difficulties, such as indefiniteness and discrete spectra of physical quantities, active measurement, stochastic interference, entanglement are all due to the appearance of incompatibility of observables (see Appendix 1). This factor represented a useful resource in structuring and sequencing the frames.

The attributes of frames describing constructs need to capture representational shifts. Their choice depends on the individual notion and may include referent (VECTOR SUPERPOSITION and OPERATOR), the role of its constituents (VECTOR), notable instances and physical situations (SUPERPOSITION: the ability to produce interference), features shaping its structural role (VECTOR SUPERPOSITION: the existence of constraints).

Analogously to the case discussed by Schurtz and Votsis (2014), we use the same set of attributes for the two versions of the concept, although some of them are chosen from the perspective of QM. As a matter of fact, the dichotomy compatibility/incompatibility is irrelevant in CM, and while the set of values of unconstrained quantities may be various in QM, in CM it is always coincident with $\mathbb{R}$.

Values are rarely structured in terms of lexical concepts (Margolis & Laurence, 2021). Most of them describe complex conditions and properties of scientific concepts and constructs.

Further considerations are worth to be included in this discussion. As we illustrate in detail in the next section, our frame representation of the evolution of individual notions often presents an overlap of classical and quantum values (grey boxes). In our format of frames, this highlights a continuity between CM and QM, and not an anomaly as in Andersen et al. (2006). Finally, we need to mention a limitation of this description: a taxonomic analysis is not sufficient to identify or visualize changes in the network of concepts. However, the category tree of a scientific concept often contains references to other basic terms of the theory. In a sense, it is not only a description of its internal structure, but also a partial representation of the inferential network of propositions the concept participates in (Amin et al., 2014). In addition, the classical version of the network is expected to be approximatively known by students. By analyzing the frames, it is possible to pinpoint some differences between the two networks. Taking into account these two factors, a carefully designed set of frames might constitute a basis to promote the revision of the relations between the concepts under scrutiny.

The frames we present in this article are the following: SYSTEM QUANTITY, MEASUREMENT, STATE, TIME EVOLUTION, GENERAL MODEL, VECTOR, VECTOR SUPERPOSITION, OPERATOR. All but TIME EVOLUTION and OPERATOR have been used to develop a teaching/learning sequence on QM for secondary school students and computer science majors. The sequence has been partially described in Authors_b, along with some preliminary data on learning. Three patterns emerge from the frames and are described together with them in the next section.

**5.3 Purpose 3: Frame Representation of Dynamics in Concepts and Mathematical Constructs**

The preliminary discussion of some general aspects can guide us in the interpretation of specific frames. First, the order of presentation has been chosen to linearize the external relations between intertwined concepts and constructs: each frame includes external references only to previously presented frames. The first one, SYSTEM QUANTITIES, introduces the basic notion of incompatibility and does not display external references to other frames. Second, frames include two types of attributes: those in which values are mutually exclusive, and those in which a sub-concept can take more than one value at the same time. In this respect, we follow the examples of Andersen et al. (2006) and Schurtz and Votsis (2014). Finally, while attributes are not all logically independent, in most cases their relationships are not displayed in the frames.

The first frame deals with changes in the concept of SYSTEM QUANTITY (Fig. 4). This expression refers to quantities describing properties of systems and includes both dynamical variables, that in QM became observables, and parameters such as charge and mass, that in non-relativistic QM behave as classical quantities. As it is possible to see by inspecting the colors of the values, all of them belong either to QM (black) or to both theories (gray), which means that the classical version of the concept is a categorical subset of the quantum one. This allows us to identify the first pattern of change: *categorical generalization*, that applies also to MEASUREMENT and STATE (Fig. 5 and 6).

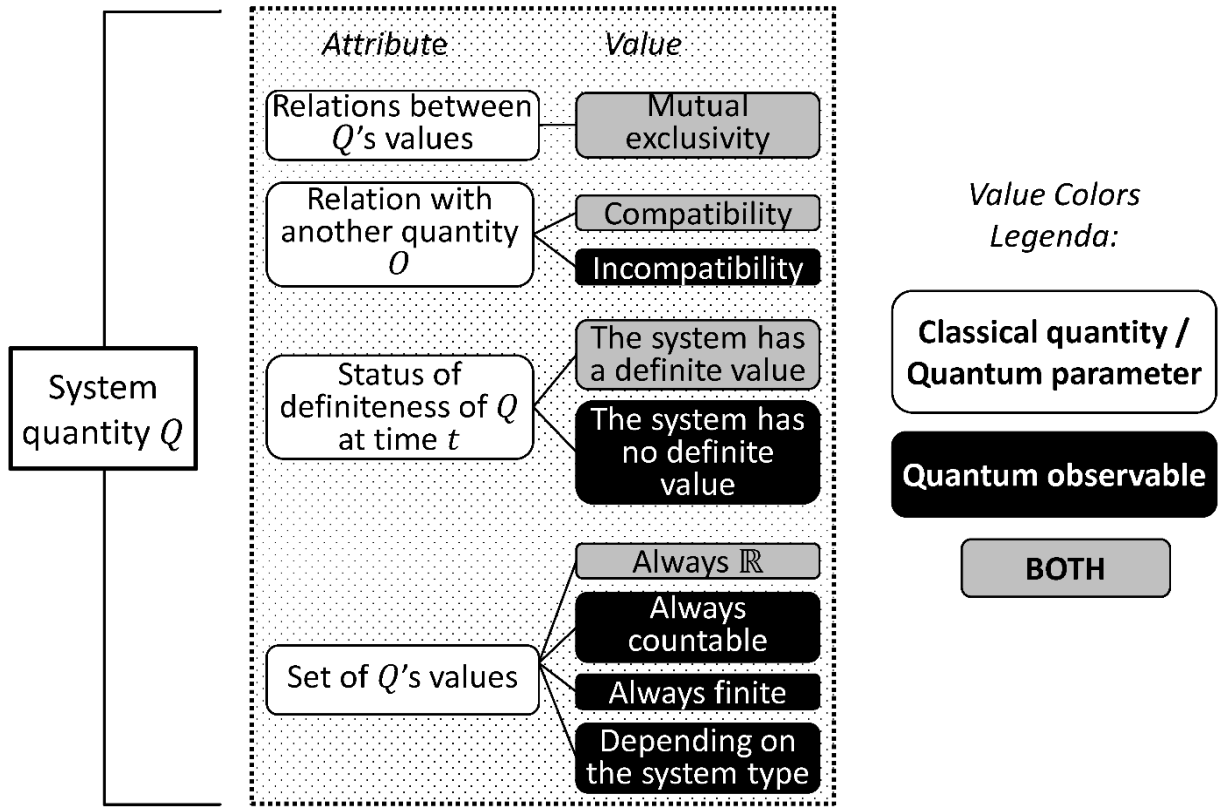

**Fig. 4** SYSTEM QUANTITY

It is important to emphasize that by this expression we do not mean that each instance of the quantum version of the superordinate concept (e.g., position, momentum) can be seen as a generalization of its classical version in a mathematical or logical sense, but that generally, the cognitive evolution required in learning this class of concepts in QM is a specific kind of conceptual revision.

As we said, MEASUREMENT and STATE frames follow the same pattern. The first one describes only ideal measurements in CM and QM, and for this reason includes only quantum uncertainty (stochastic acquisition of a value), but not uncertainty due to external factors or to the limited sensibility of an experimental setup.

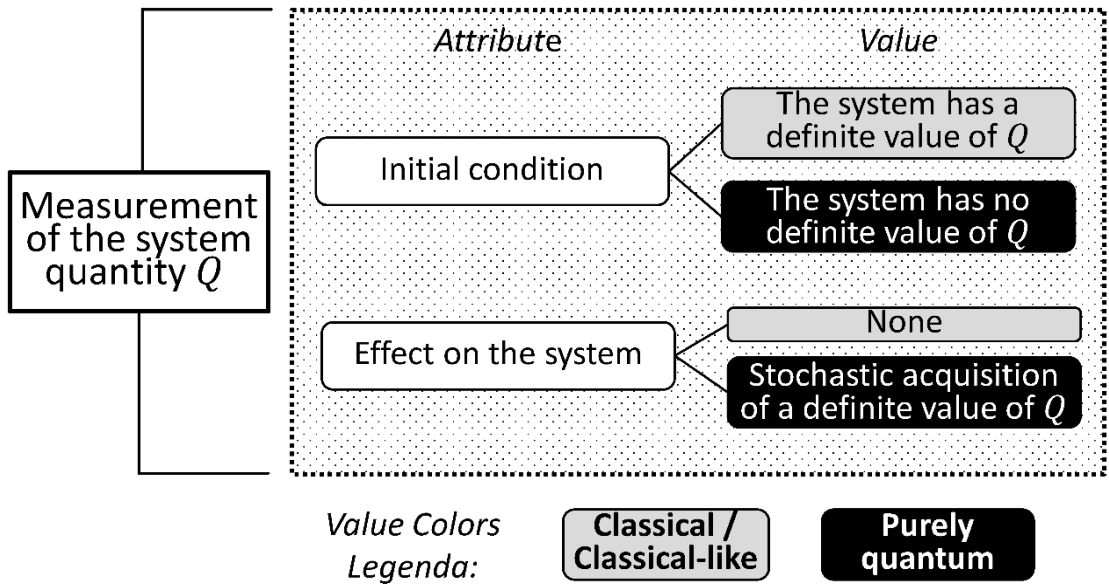

Fig. 5 MEASUREMENT

The examples of classical state displayed in the frame are those upper undergraduate students and secondary school students may be familiar with: respectively states of Hamiltonian mechanics and the thermodynamic state (see Fig. 6). The attribute PHYSICAL INFORMATION is connected to the attribute TYPE OF INFORMATION by an arrow to indicate that the latter describes an educationally significant aspect of the former. In fact, the values of the second attribute can be logically derived from the values of the first.

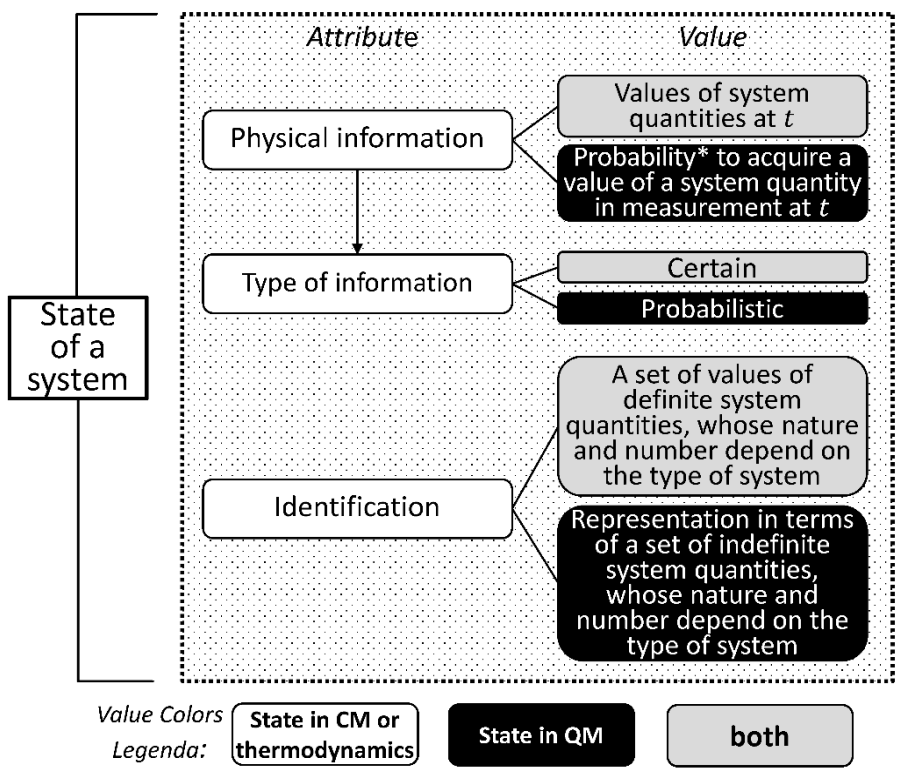

* The term "probability" is to be intended in the interval [0,1[.
  If probability is 1, the system already has a definite value at time $t$

**Fig. 6**   STATE

Fig. 7 depicts the structure of TIME EVOLUTION in the absence of measurement. Since the classical counterpart of elementary quantum systems is a point-like particle, a null torque was not included among the classical conditions of physical invariance. Here the distribution of colors displays a more varied picture, dominated by another pattern: *value disjunction*. This expression means that attributes generate only a white value and a black value, thus indicating that the classical and the quantum versions of the concept are completely different as regards the attributes involved. The radical nature of this change is mitigated by the similarity between Hamiltonian mechanics and QM with respect to the general meaning of time variance and invariance (the type of change is depicted by grey values). However, the theory-dependent nature of these forms of evolution can be identified by referring to the concept of STATE and to its frame.

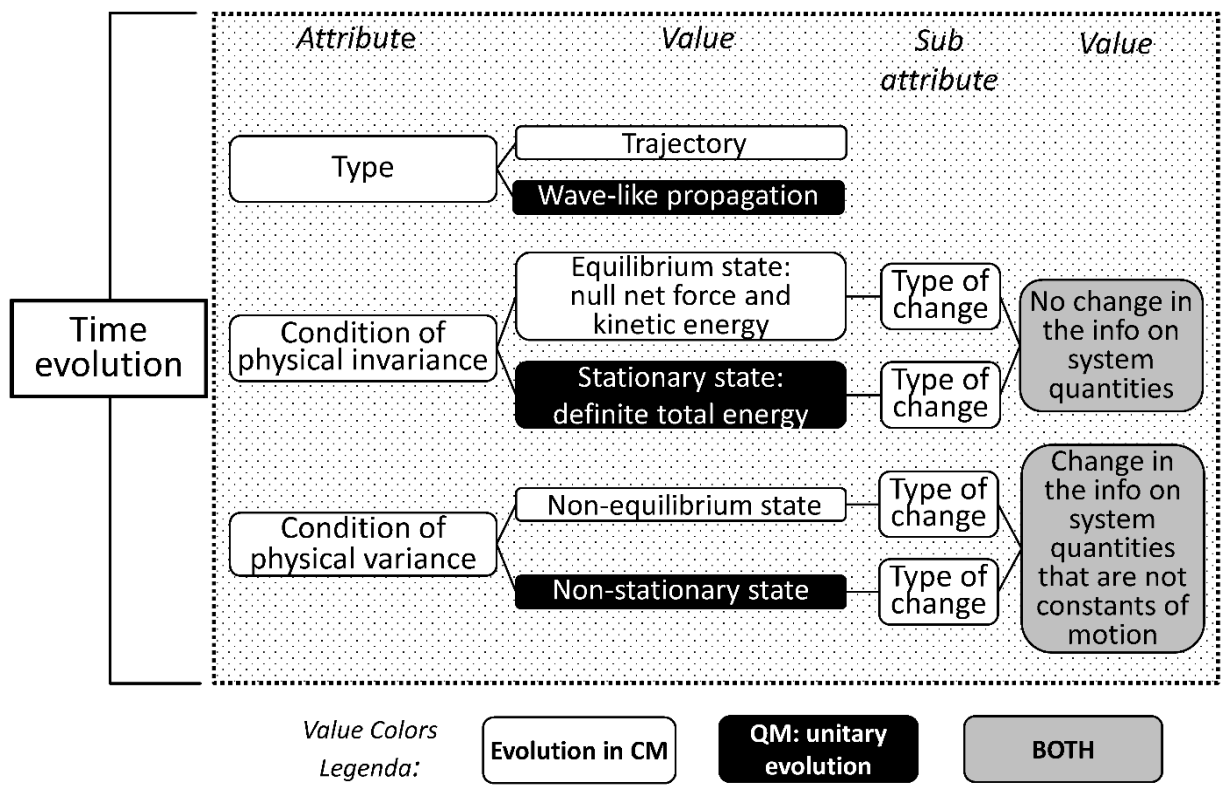

**Fig. 7**   TIME EVOLUTION

The GENERAL MODEL is described by means of two frames (Fig. 8 and 9). The first concerns the most basic aspects of the wave-particle duality and is discussed in terms of two processes: DETECTION and PROPAGATION. Attributes have been selected by adopting an experimental operative perspective. Detection and Identification of the which-path information are treated as performed by ideal devices. This frame reveals the third and last pattern, *change in value constraints*: by changing the constraints between values belonging to the classical wave model and particle model, we move on to the quantum model (grey values), which reinterprets aspects of both.

**Fig. 8** GENERAL MODEL – WAVE-PARTICLE DUALITY

However, when we look deeper and need to assess how we detect the signature of the wave nature of propagation (interference), the picture changes completely: the theoretical shift is dominated by *value disjunction* (Fig. 9).

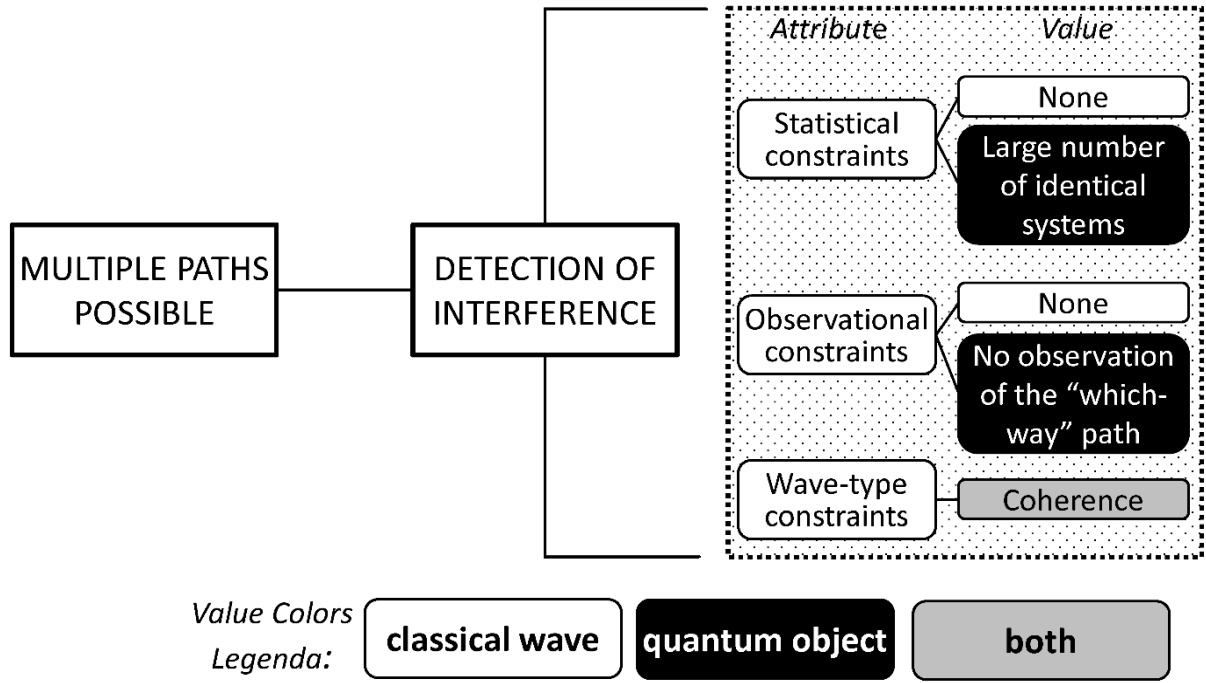

**Fig. 9** GENERAL MODEL – STOCHASTIC INTERFERENCE

Moving on to examine frames describing constructs like VECTOR and VECTOR SUPERPOSITION (Fig. 10 and 11), we observe that the representational shift is stronger: as regards referent, representational and conventional properties, notable physical situations described by the constructs (only depicted in the frame on VECTOR SUPERPOSITION), and other features that affect their structural role, *value disjunction* is the dominant pattern. The old and new version have little in common. In the frame on vectors, we included the value "vectors defined up to a phase", since at high school level we can discuss this issue in the real state spaces of linear polarization and spin-x and -z, restricting the statement to "vectors defined up to a sign",

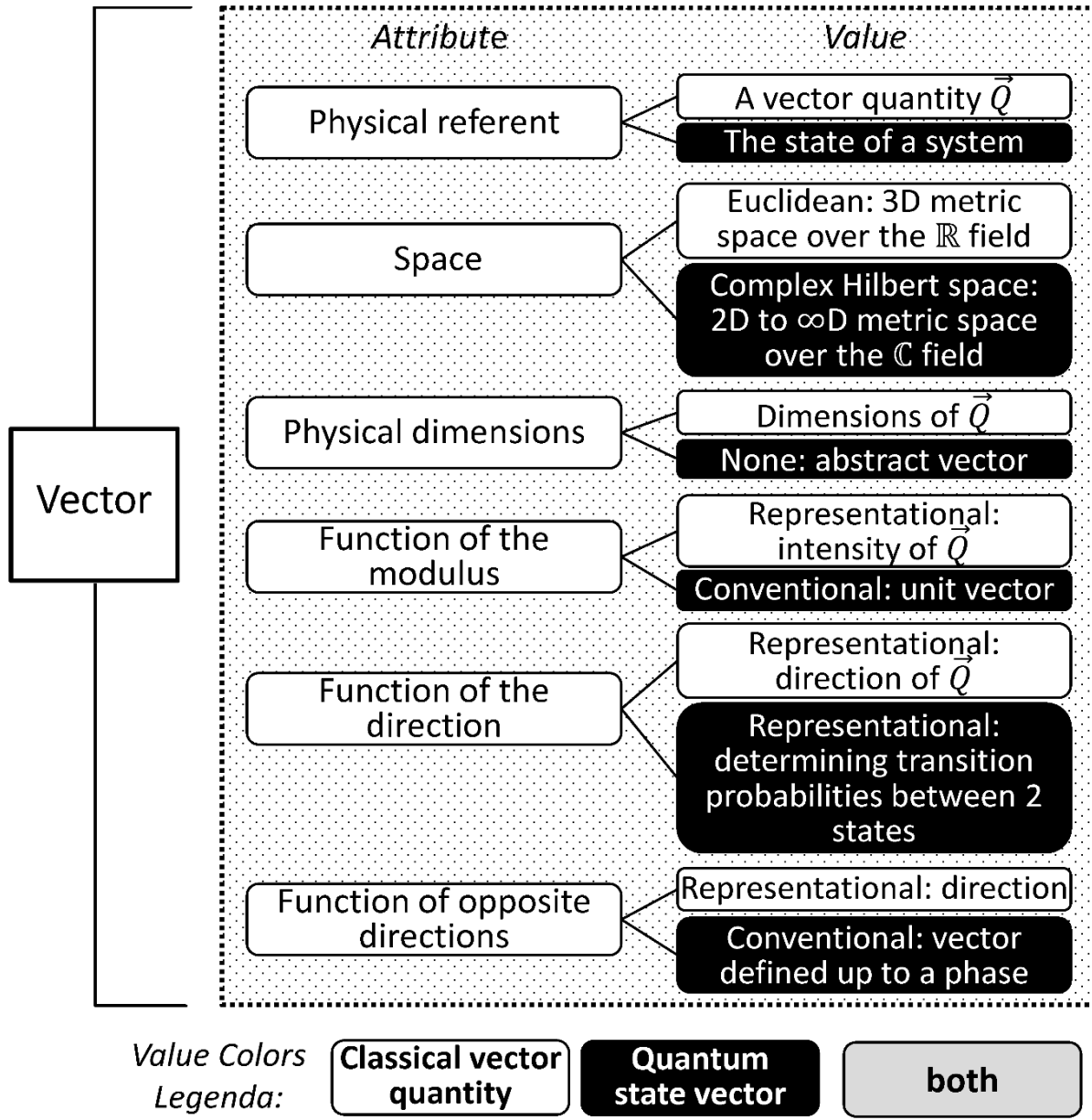

**Fig. 10** VECTOR

In structuring the VECTOR SUPERPOSITION frame, we did not compare two version of the mathematical process, but three: two are classical (superposition of forces and of waves) and one quantum. The color code we use to depict values changes: here white values belong to superposition of forces, gray ones to superposition of waves, black to quantum superposition. When two of these sub-concepts share the same value, we paint its box with both colors. Except for the number of physical entities involved, *value disjunction* is the pattern displayed by the frame.

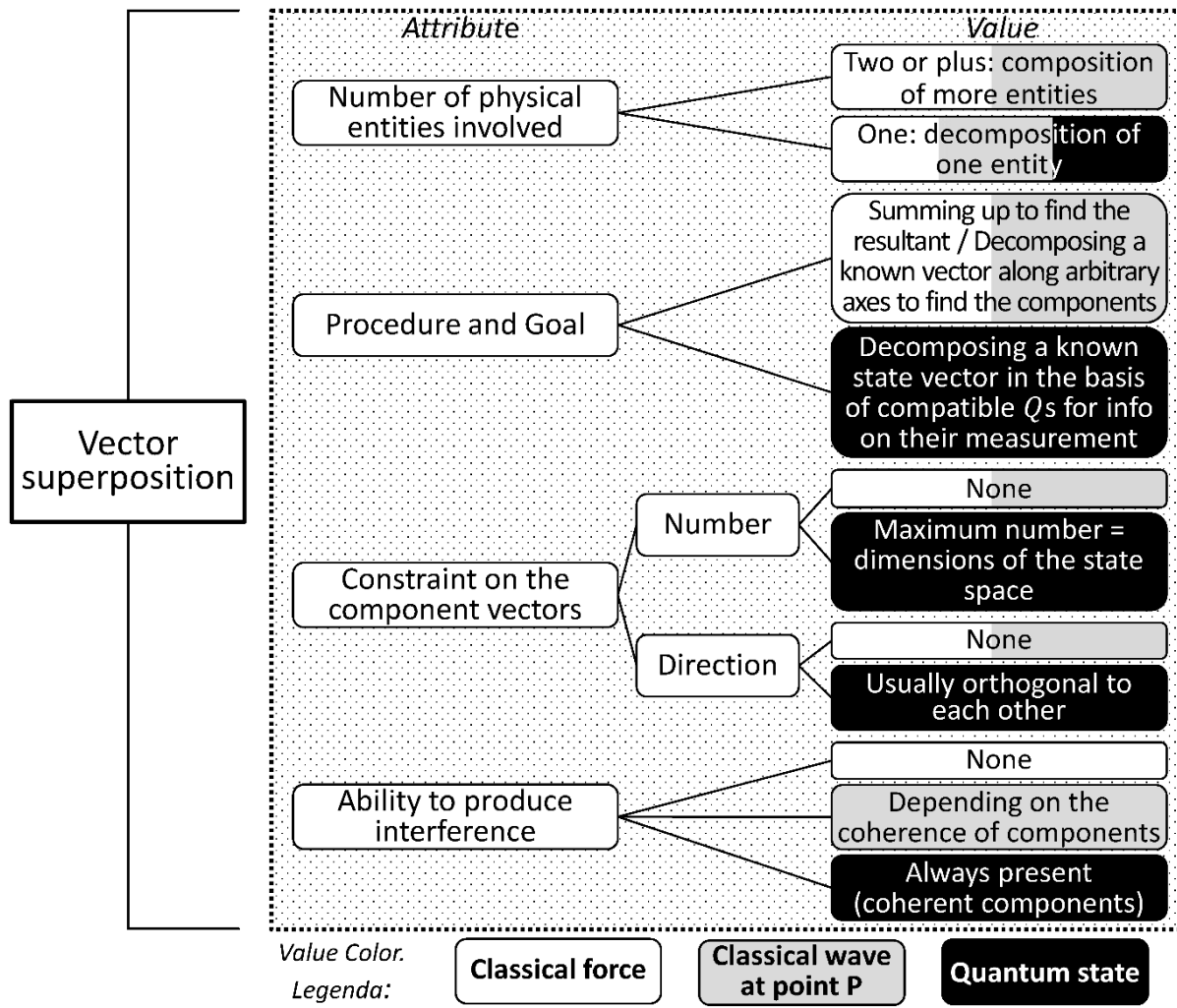

**Fig. 11** VECTOR SUPERPOSITION

The general structure of the OPERATOR frame (Fig. 12) is similar to that used for superposition: we compared three different forms of operators. This time one is classical and two are quantum. We refer to operators in Newtonian mechanics (white values), quantum unitary operators (grey), and quantum Hermitian operators (black). As we see, the only similarity between a unitary and a Hermitian operator is the vector space in which they are defined. However, for our purposes, what matters is that Hermitian operators and Newtonian ones are totally different entities (*value disjunction*).

**Fig. 12** OPERATOR

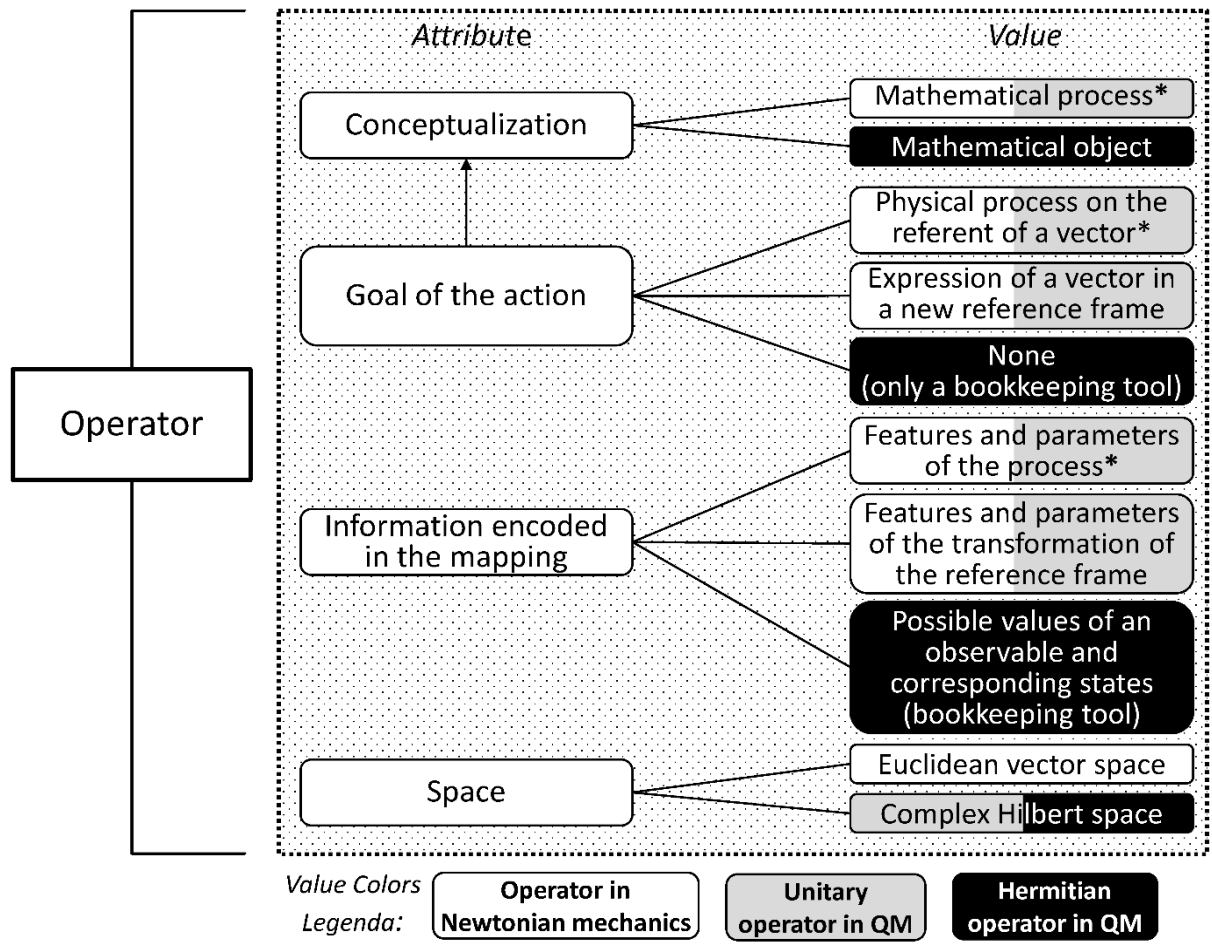

* A unique instance of Hermitian operator is associated with this value: the Hamiltonian operator, related to time evolution

## 6 Purpose 4: Pattern-Dependent Suggestions on a Productive Use of Prior Intuition

Visualizing categorical continuity and change by means of dynamic frames may represent a support in addressing fundamental issues in the learning of successive theories, especially if – as is the case of QM – the new theoretical picture involves a clear-cut detachment from more intuitive frameworks and the loss of visualization of its referents in the world.

First, the frame representation of continuity and change in concepts and constructs allow us to draw and visualize a comparison of basic notions of the older and the new paradigm, not in terms of limiting processes (as in the case of relativity), but of categorical structure. In addition, the patterns in the trajectories of these entities may help us identify strategies to put prior intuition in the service of learning QM and may work as a guide to curricular design. As we will see in the following sections, the most important indications we can draw in this sense concern the possible role of prior cognitive structures in the building of concepts of the new theory: the availability of such information poses tight constraints on the types of educational strategies which are likely to be productive. Furthermore, in all cases the dynamic frame representation can be used directly to build activities in the form of comparison tables between an old and a new version of a concept or construct, to be completed by students, thus activating metacognitive reflection.

The rest of this section is our implementation of Purpose 4. Examples are taken from the development of a teaching/learning sequence on QM for secondary school students and computer science majors (Authors_b).

**6.1 Categorical Generalization**

For notions evolving in terms of categorical generalization, we can rely on student conceptual resources concerning the old version of the concept to discuss those instances of the new one that are common to both theories. Then, the new features of the concept can be made available by adopting constructivist strategies. Finally, according to the prevalence model of CC (Potvin et al., 2015), we can use the relevant frame to structure end of unit tables containing interpretive tasks that allow students to discriminate between the older and the new values of the superordinate notion, identifying the correct context of application of each one.

An example of the suggested strategy is the introduction of the concept of quantum measurement in the context of polarization. We found that many students could not reconcile the classical description of polarization measurements with the quantum one. Classical polarization can have any orientation and is identified by measuring its angle. The polarization of a photon can also have any orientation, but its measurement gives one of two angles that may be different from the initial one. In particular, students were reluctant to accept the purely quantum feature of measurement, i.e., the idea that it can involve a value-altering interaction (see corresponding frame in Fig. 5). To overcome this difficulty, we structured the following activity: students were asked to interpret a situation in which a random mixture of photons with horizontal (0°) or vertical polarization (90°) passes a birefringent crystal with rays at 0° and 90°, one photon at a time, and is collected by one of two detectors positioned in the two rays. Since the angle of polarization of each photon is revealed without altering it, most students consistently interpreted the interaction as a measurement by using their classical intuition. After that, we introduced quantum measurement as a generalization of the classical one without finding any resistance: a polarization measurement in QM is the interaction of a photon with the aforementioned device, independently from its initial polarization. The nature of this behavior was explained by introducing the new quantum feature: *incompatibility*. In order to build the missing schema, we had students explore photon polarization measurements at both phenomenal and mathematical level. Subsequently, we asked them to identify in what conditions there can be a change in the polarization of a photon in measurement. Two possible relations between polarization angles were found by students, corresponding to *unaquirability* (if a system has one value, it cannot possess the other as well, or acquire it in a measurement of the corresponding observable – true for perpendicular directions) and, as hoped for, *incompatibility* (if the system has one value, it loses it and may stochastically acquire the other in the measurement of the corresponding observable – true for all other couples of directions). The instructors justified the new relation accordingly, as an empirical regularity: in QM, experimental observations show that values of some quantities are incompatible with each other (by acquiring one in measurement, the system loses the other).

**6.2 Value Disjunction**

In this case, the new version of the notion is radically different from the old one. The main difference between the strategy suggested for categorical generalization and the strategy we recommend here, is that we need to start with the construction of the new features (e.g., phenomenal inquiry for scientific concepts, mathematical sense-making and embodied cognition for mathematical constructs) and use prior intuition as a contrast (Henderson et al., 2017). Although many authors in the past have used cognitive conflict strategies in the initial stages of the teaching-learning process, in order to first get rid of old conceptions and then introduce new ones, in the perspective of the prevalence model of CC (Potin et al., 2015), prior intuition can be contrasted with new features at the end of the instructional sequence: the frame representation of the notion can be used to structure interpretive tasks with the aim to install inhibitory stop signs. The purpose of these signs would be to identify the old features that lead to unproductive reasoning in the new theoretical context.

In the case of classical and quantum mechanics, this pattern is primarily associated with mathematical constructs (Fig. 10, 11, 12). The impossibility to visualize quantum systems and the unfamiliar nature of the new version of the concepts represent an educational bottleneck that can be overcome also with the support of mathematics. However, analogies in mathematical constructs used in both theories are deceptive and may lead to interpretive pitfalls. The case of quantum superposition is paradigmatic: while prior knowledge of its classical version can be productive for identifying the role of the coefficients in simple contexts such as polarization (their square stands for transition probability), the change in referent is a totally different matter. As we saw in Section 4.2, even graduate students struggle to build a consistent understanding of the new referent (goal and procedure of the mathematical process). In this task, embodied cognition can be of help, providing the grounds for building a mental simulation of an abstract notion (Zohar et al., 2021). According to Redish and Kuo (2015), p. 570: "our understanding of many mathematical concepts relies on everyday ideas such as spatial orientations, groupings, bodily motion, physical containment, and object manipulations (such as rotating and stretching)." In the context of polarization, we made use of the perceptual experience of passive rotations (Fig. 13).

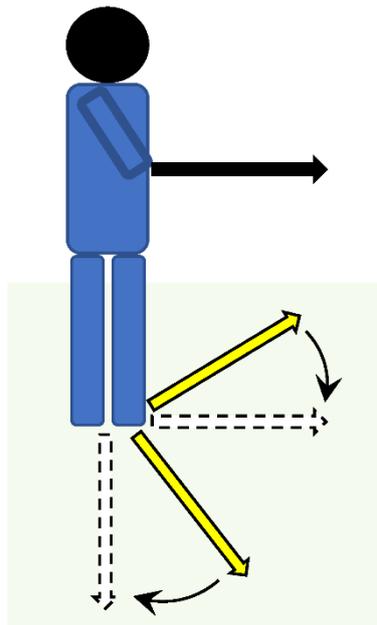

**Fig. 13** Embodying the referent of quantum superposition

A student points an arrow of a given length at a fixed angle on the horizontal plane (the state vector). Two arrows of the same length departing from the center are drawn on the ground to identify perpendicular directions (eigenvectors of the observable). Rotating the direction of the arrows corresponds to changing the observable under scrutiny. At qualitative level, when the direction of one arrow on the ground coincides with the direction of the arrow held by the student, it means that the chosen observable has a definite value, otherwise it has none. Embodied models like this one may be useful to promote a correct interpretation the conceptual referent of quantum superposition: a decomposition of one physical entity (the vector representing the state of the system) in the eigenbasis of a given observable to obtain information on its measurement, as opposed to the most typical classical situation: a composition of more physical entities (vectors representing different interactions or waves) with the aim to find the resultant. As it happens in the decomposition of a vector, the resultant is known from the start. The frames on vectors (Fig. 10) and vector superposition (Fig. 11) were used at the end to structure a table including interpretive tasks on notable differences between the features of superposition of forces, of waves, and of quantum state vectors, and between the concepts described by these vectors, thus using prior intuition as a contrast.

**Fig. 14** Quantum superposition: activities in the form of comparison tables

### 6.3 Change in Value Constraints

This pattern corresponds to the appearance of a new instance of a concept (new subordinate concept) whose properties are a reorganization of the properties belonging to old instances. In learning, the accomplishment of this change appears to be the simplest, since all the relevant properties are already available to students, and the new instance is a coalescence of individual constituents of different old instances (Carey, 1999). However, the old subordinate concepts may be ontologically very different from each other, and in this case the introduction of a bridging analogy may be needed to promote the reconfiguration of existing resources. Evidence has been provided on the fact that the use of analogies is productive especially with older students, who have the prior knowledge required to understand how the analogy could be correctly mapped to explain new phenomena (Vosniadou & Skopeliti, 2017). Exactly the kind of population that could be exposed to the learning of a successor of a scientific theory.

In the transition from CM to QM, this pattern concerns the most basic aspects of the wave-particle duality (Fig. 8). Unfortunately, in order to fully interpret the wave nature of propagation, students must develop an understanding of quantum interference (Fig. 9), that is dominated by value disjunction. Therefore, a combination of strategies related to both patterns is needed. In QM, interference is a physical consequence of incompatibility of observables and of the use of superposition to describe measurements on a system. As a result, we decided to address first the vector representation of a state, the superposition of state vectors, and its relation with particle detection (measurement), as a prerequisite to discussing the propagation of a system in a "which-way" experiment and to introduce interference for substantiating its wave nature. Value disjunction is clearly visible in the following features of the quantum version of interference: it is totally internal to the individual system – from which the counterintuitive statement: "Each photon then interferes only with itself" (Dirac, 1967, p. 9) - but turns into an observable process only by using a large number of identically prepared systems, and as long as no observation of the "which-way" path is performed (see Fig. 9). When students have developed a consistent understanding of these aspects, we propose the concept of field as a bridging analogy between the particle behavior (in measurement) and the wave behavior (in the propagation) of a quantum system. Similarly to the classical electromagnetic field, this field supports the wave associated with the system in vacuum and allows students to interpret how the system knows whether two paths are open or only one of them. Differently from classical fields, its interaction with matter is punctual in space (we can identify the detector where it takes place) but involves the entire spatially extended field in a non-local way (it instantaneously disappears). These new features had been made available to students in advance, during the discussion of measurement and of interference.

**7 Conclusion and Future Directions**

In this work, we developed an initial proposal for modeling CC in learning a successor of a theory students are familiar with. Its generative potential for research and design was explored by applying the model to a concrete case in order to provide a structural account of the educational transition between the theories under scrutiny, identify new types of challenges that are absent in the learning of science at introductory level, develop representational tools to support concept building, and suggest ways to leverage the use of prior intuition for developing an understanding of the new theory.

The most significant differences between this kind of CC and the one familiar to the science education community concern the cognitive factors, namely the challenges connected with changes at different representational levels and in exemplars. With relation to theory change, epistemic cognition represents a deep and open issue that needs further investigation. Metacognition, affective and sociocultural factors, instead, do not appear to show significant differences from those described in literature, except for aspects connected with epistemology (see Section 2.2). This does not mean that such factors can be ignored in the design of educational environments for the learning of successive theories, but that they should be considered according to the educational context, independently of the theory change at hand.

While the framework for describing the different types of change that affect each representational level has been developed by using as a reference the classical-quantum transition in physics, its basic features may be applicable to different disciplines. The same can be said for the techniques introduced for connecting theory change to cognitive challenges and exploring individual details of these challenges. Since many changes at all levels give rise to learning difficulties, this kind of exploration can be used to elicit potential issues in need of investigation.

The frame representation of categorical change has shown its productivity in allowing us to draw a clear comparison of basic notions of the older and the new paradigm in terms of categorical structure, and in revealing recognizable patterns: categorical generalization, value disjunction and change in value constraints. Pattern-dependent strategies have been suggested to put prior intuition in the service of learning the new theory and as a guide to curricular design. Also in this case, the applicability of the tool may go beyond the specific theory change under scrutiny.

Future directions include, in the first place, empirically testing the productivity of the model by evaluating the strategies stemming from it in teaching experiments on large samples of students. Second, applying the model to other domains involved in the second quantum revolution and extending it for exploring change in exemplars and epistemological change. As regards exemplars, their context dependence suggests a "knowledge in pieces" approach (diSessa, 2014). Another line of research could be the design of pre-post tests on the interpretation and use of the relevant concepts involved in theory change. In this context, frames could be used in the development of a rubric for the test. A further possibility is evaluating the productivity of the model in the learning of successive theories in disciplines whose conceptual, mathematical, and epistemological nature is very far from physics. An example could be the change from a microeconomic theory as developed within the neoclassical framework to behavioral microeconomics (Dhami, 2016, pp. 1-21).

**Appendix 1. The Fundamental Role of Incompatibility of Observables**

According to Masterton et al. (2016), the theory change from classical mechanics (CM) to quantum mechanics (QM) is due to the appearance of incompatibility of observables. In mathematical terms, a change in the algebra of observables. In order to specify how this feature affects the structure and the sequence of our frame representation, we briefly describe the main physical consequences of the removal of incompatibility from the theoretical framework of QM. Masterton et al. (2016) drew their conclusions by working in the Wigner formulation of QM. Since we're writing for educational purposes, we follow a different path: we adopt the Schrödinger picture of the Hilbert space formulation, which is currently used in instruction, and perform a counterfactual experiment by answering the question: "what if all observables corresponding to classical dynamical variables were compatible?". In mathematical terms, that is to say: "what if $[\hat{x}, \hat{p}] = 0$ ?".

In Hamiltonian mechanics, dynamical variables are those quantities that can be mathematically represented by functions of $x$ and $p$. In the transition to QM, they become observables, represented by operators. However, there is another class of observables: all the abstract observables that can defined by selecting an arbitrary decomposition of the Hilbert space and associating a value to each subspace. Given that dynamical variables are the only physically measurable class of observables, we restrict the discussion to them

*Disappearance of discrete spectra*: the number of values some observables can take is limited by the constraint between the position operator and the momentum operator ($[\hat{x}, \hat{p}] = i\hbar$). Take, for instance, the case of angular momentum: if $[\hat{x}, \hat{p}] = 0$, all of the commutators between different components of the angular momentum would be zero, and therefore no ladder operator would arise to impose the construction of a discrete spectrum. The discrete nature of some quantities was already accounted for in the old quantum theory. In this respect, QM differs inasmuch as it explains the existence of discrete spectra as a consequence of incompatibility.

*Superposition states are indistinguishable from statistical mixtures*: since all dynamical variables are compatible, there exists a complete basis of common eigenstates that have definite values of all the variables. The results of measurement of every dynamical variable on an arbitrary superposition of these states are indistinguishable from those performed on the diagonal mixture composed of the same states and probabilities. No interference may arise in measurement.

*All correlations in measurement can be interpreted by means of a probabilistic CM*: with the removal of incompatibility, quantum entanglement disappears too. This can be derived from the fact that the violation of Bell's inequalities can be theoretically predicted only by means of measurements on incompatible observables.

*But time evolution in the absence of measurement does not reduce to the classical one*: some authors prefer to interpret our counterfactual experiment as a limiting process on the Planck constant: $\hbar \to 0$ (e.g., Klein, 2012). According to Klein's line of reasoning, we obtain a probabilistic theory that contains a much larger number of states not belonging to CM and recovers Newtonian predictions only as a limiting case and under given conditions. Almost all states do not admit a transition from QM to CM in the limit $\hbar \to 0$.

Incompatibility of observables is the most remarkable new property of quantum concepts, since it is an essential condition behind the appearance of indefiniteness and discrete spectra in physical quantities, active measurement, stochastic interference, and entangled correlation in measurements on composite systems. Therefore, we choose to display it at the very beginning of our frame representation, which starts with the frame on SYSTEM QUANTITIES.

**Conflict of Interest Statement**

The authors have no conflicts of interest to declare that are relevant to the content of this article. All authors certify that they have no affiliations with or involvement in any organization or entity with any financial interest or non-financial interest in the subject matter or materials discussed in this manuscript. The authors have no financial or proprietary interests in any material discussed in this article.

**References**

Amin, T.G., Smith, C., & Wiser, M. (2014). Student conceptions and conceptual change: Three overlapping phases of research. In N. G. Lederman & S. K. Abell (Eds.), *Handbook of research on science education, Volume II*, (pp. 71-95). Routledge. https://doi.org/10.4324/9780203097267-12

Amin, T.G. (2015). Conceptual metaphor and the study of conceptual change: Research synthesis and future directions. *International Journal of Science Education*, 37(5-6), 966-991. https://doi.org/10.1080/09500693.2015.1025313

Amin, T. G., & Levrini, O. (2017). Overall synthesis: Facing the challenges of programmatic research on conceptual change. In T. G. Amin, & O. Levrini. (Eds.), *Converging perspectives on conceptual change: Mapping an emerging paradigm in the learning sciences* (pp. 334-351). Routledge. https://doi.org/10.4324/9781315467139-38

Amin, T. G., (2017), Synthesis II: Representations, concepts and concept learning. In T. G. Amin & O. Levrini. (Eds.), *Converging perspectives on conceptual change: Mapping an emerging paradigm in the learning sciences* (pp. 129-149). Routledge. https://doi.org/10.4324/9781315467139

Andersen, H., Barker, P., & Chen, X. (2006). *The cognitive structure of scientific revolutions*. Cambridge University Press. https://doi.org/10.1017/CBO9780511498404

Arabatzis, T. (2020). What are scientific concepts? In K. McCain, & K. Kampourakis (Eds.) *What is scientific knowledge? An introduction to Contemporary Philosophy of Science* (pp. 85-99). Routledge. https://doi.org/10.4324/9780203703809-6

Authors_a

Authors_b

Authors_c

Authors_d

Authors_e

Ayene, M., Kriek, J., & Baylie, D. (2011). Wave-particle duality and uncertainty principle: Phenomenographic categories of description of tertiary physics students' depictions. *Physical Review Special Topics - Physics Education Research*, 7(2), 1–14. https://doi.org/10.1103/PhysRevSTPER.7.020113

Baily, C., & Finkelstein, N. D. (2010). Refined characterization of student perspectives on quantum physics. *Physical Review Special Topics - Physics Education Research*, 6(2), 1–11. https://doi.org/10.1103/physrevstper.6.020113

Ballentine, L. E. (2014). *Quantum mechanics: A modern development. 2$^{nd}$ Edition*. World Scientific. https://doi.org/10.1142/9038

Belloni, M., Christian, W., & Brown, D. (2007). Open source physics curricular material for quantum mechanics. *Computing in Science & Engineering*, 9 (4), 24-31. https://doi.org/10.1109/mcse.2007.80

Brookes, D. T., & Etkina, E. (2007). Using conceptual metaphor and functional grammar to explore how language used in physics affects student learning. *Physical Review Special Topics - Physics Education Research*, 3(1), 1–16. https://doi.org/10.1103/PhysRevSTPER.3.010105

Bub, J. (1997). *Interpreting the quantum world*. Cambridge University Press.

Carey, S. (1999). Sources of Conceptual Change. In E. K. Scholnick, K. Nelson, S. A. Gelman, P. H. Miller (eds.), *Conceptual development: Piaget's legacy* (pp. 293-327). Lawrence Erlbaum Associates Publishers.

Carey, S. (2000). Science education as conceptual change. *Journal of Applied Developmental Psychology*, 21 (1), 13-19. Psychology Press. https://doi.org/10.1016/S0193-3973(99)00046-5

Chi, M. T. H. (2008). Three types of conceptual change: Belief revision, mental model transformation, and categorical shift. In Vosniadou, S. (ed.) *International handbook of research on conceptual change (1$^{st}$ ed.)*. pp. 61-82. Routledge. https://doi.org/10.4324/9780203874813

Coppola, P., & Krajcik, J. (2013). Discipline-centered post-secondary science education research: Understanding university level science learning. Journal of Research in Science Teaching, 50(6), 627–638. https://doi.org/10.1002/tea.21099

Dhami, S. S. (2016). *The foundations of behavioral economic analysis*. Oxford University Press.

Dieks, D., & Vermaas, P. E. (1998). *The modal interpretation of quantum mechanics (Vol. 60)*. Springer Science & Business Media.

Dini, V., & Hammer, D. (2017). Case study of a successful learner's epistemological framings of quantum mechanics. *Physical Review Physics Education Research*, 13(1), 1-16 https://doi.org/10.1103/PhysRevPhysEducRes.13.010124

Dirac, P. A. M. (1967). *The principles of quantum mechanics, 4th ed. (revised)*. Clarendon Press.

diSessa, A. A. (2014). A history of conceptual change research: Threads and fault lines. In R. K. Sawyer (Ed.), *The Cambridge handbook of the learning sciences, Second edition* (pp. 88-108). Cambridge University Press. https://doi.org/10.1017/CBO9781139519526.007

diSessa, A. A., & Sherin, B. L. (1998). What changes in conceptual change? *International Journal of Science Education*, 20 (10), 1155-1191. https://doi.org/10.1080/0950069980201002

diSessa A. A., Sherin, B. L., Levin, M. (2016). Knowledge analysis: An introduction. In A. A. diSessa, M. Levin, & N. J. S. Brown (Eds.), *Knowledge and interaction: A synthetic agenda for the learning sciences* (pp. 30-71). Routledge. https://doi.org/10.4324/9781315757360

Dubson, M., Goldhaber, S., Pollock, S., & Perkins, K. K. (2009). Faculty disagreement about the teaching of quantum mechanics. In *Physics Education Research Conference 2009 Proceedings* (pp. 137–140). American Institute of Physics. https://doi.org/10.1063/1.3266697

Elby, A., & Hammer, D. (2010). Epistemological resources and framing: A cognitive framework for helping teachers interpret and respond to their students' epistemologies. In L. D. Bendixen & F. C. Feucht (Eds*.), Personal epistemology in the classroom: Theory, research, and implications for practice* (pp. 409–434). Cambridge University Press. https://doi.org/10.1017/CBO9780511691904.013

European Quantum Flagship (2020, February). *Strategic Research Agenda*. https://ec.europa.eu/newsroom/dae/document.cfm?doc_id=65402

Gamerschlag, T., Gerland, D., Osswald, R., & Petersen, W. (2014). General introduction. In T. Gamerschlag, D. Gerland, R. Osswald, & W. Petersen (Eds.), *Frames and concept types: Applications in language and philosophy* (pp. 3-21). Springer. https://doi.org/10.1007/978-3-319-01541-5

Gärdenfors, P. (2000). *Conceptual Spaces. The Geometry of Thought*. The MIT Press. https://doi.org/10.7551/mitpress/2076.001.0001

Gire, E., & Manogue, C. A. (2011). Making sense of quantum operators, eigenstates and quantum measurements. In *Physics Education Research Conference 2011 Proceedings* (pp. 195–198). American Institute of Physics. https://doi.org/10.1063/1.3680028

Goldhaber, S., Pollock, S., Dubson, M., Beale, P., & Perkins, K. (2009). Transforming upper-division quantum mechanics: Learning goals and assessment. In *Physics Education Research Conference 2009 Proceedings* (pp. 145-148). American Institute of Physics. https://doi.org/10.1063/1.3266699

Griffiths, R. B. (2001). *Consistent quantum theory*. Cambridge University Press. https://doi.org/10.1017/CBO9780511606052

Henderson, J. B., Langbeheim, E., & Chi, M. T. (2017). Addressing robust misconceptions through the ontological distinction between sequential and emergent processes. In T. G. Amin, & O. Levrini, (Eds.), *Converging perspectives on conceptual change: Mapping an emerging paradigm in the learning sciences* (pp. 26-33). Routledge. https://doi.org/10.4324/9781315467139-5

Hofer, B. K., & Bendixen, L. D. (2012) Personal epistemology: theory, research and future directions. In K. R. Harris, S. Graham, T. Urdan, C. B. McCormick, G. M. Sinatra, & J. Sweller (Eds.), *APA educational psychology handbook, Vol. 1: Theories, constructs, and critical issues* (pp. 227–256). American Psychological Association. https://doi.org/10.1037/13273-009

Home, D., & Whitaker, M. A. B. (1992). Ensemble interpretations of quantum mechanics: A modern perspective. North-Holland. https://doi.org/10.1016/0370-1573(92)90088-H

House of Representatives 6227–National Quantum Initiative Act, 115th Congress, 164:132 STAT. 5092, https://www.congress.gov/bill/115th-congress/house-bill/6227, 2018.

Hoyningen-Huene, P. (1993). *Reconstructing scientific revolutions: Thomas S. Kuhn's philosophy of science*. University of Chicago Press.

Klein, U. (2012). What is the limit $\hbar \to 0$ of quantum theory? *American Journal of Physics*, 80(11), 1009-1016. https://doi.org/10.1119/1.4751274

Kohnle, A., Baily, C., Campbell, A. Korolkova, N., & Paetkau, M. J. (2015). Enhancing student learning of two-level quantum systems with interactive simulations. *American Journal of Physics*, 83(6) 560-566. https://doi.org/10.1119/1.4913786

Kohnle, A., & Deffenbach, E. (2015). Investigating student understanding of quantum entanglement. In *Physics Education Research Conference 2015 Proceedings* (pp. 171–174). American Institute of Physics. https://doi.org/10.1119/perc.2015.pr.038

Krijtenburg-Lewerissa, K., Pol, H. J., Brinkman, A., & Van Joolingen, W. R. (2017). Insights into teaching quantum mechanics in secondary and lower undergraduate education. *Physical Review Physics Education Research*, 13(1) 1-21. https://doi.org/10.1103/PhysRevPhysEducRes.13.010109

Krijtenburg-Lewerissa, K., Pol, H. J., Brinkman, A., & Van Joolingen, W. R. (2018). Key topics for quantum mechanics at secondary schools: a Delphi study into expert opinions. *International Journal of Science Education*, 41(3), 349-366. https://doi.org/10.1080/09500693.2018.1550273

Kuhn (1962). *The Structure of Scientific Revolutions*. University of Chicago Press.

Levrini, O. (2014). The role of history and philosophy in research on teaching and learning of relativity. In M. R. Matthews (Ed.) *International handbook of research in history, philosophy and science teaching* (pp. 157-181). Springer. https://doi.org/10.1007/978-94-007-7654-8_6

Levrini, O., & Fantini, P. (2013). Encountering productive forms of complexity in learning modern physics. *Science & Education*, 22(8), 1895-1910. https://doi.org/10.1007/s11191-013-9587-4

Marshman, E., & Singh, C. (2015). Framework for understanding the patterns of student difficulties in quantum mechanics. *Physical Review Special Topics – Physics Education Research*, 11(2), 1-26. https://doi.org/10.1103/physrevstper.11.020119

Margolis, E. & Stephen, L. (2021), "Concepts". In E. N. Zalta (Ed.), *The Stanford Encyclopedia of Philosophy*, https://plato.stanford.edu/archives/spr2021/entries/concepts/.

Masterton, G., Zenker, F., & Gärdenfors, P. (2016). Using conceptual spaces to exhibit conceptual continuity through scientific theory change. *European Journal for Philosophy of Science*, 7(1), 127-150. https://doi.org/10.1007/s13194-016-0149-x

McKagan, S. B., Perkins, K. K., & Wieman, C. E. (2008). A deeper look at student learning of quantum mechanics: the case of tunneling. *Physical Review Special Topics - Physics Education Research*, 4(2), 1–17. https://doi.org/10.1103/PhysRevSTPER.4.020103

Modir, B., Thompson, J. D., & Sayre, E. C. (2019). Framing difficulties in quantum mechanics. *Physical Review Physics Education Research*, 15(2), 1–18. https://doi.org/10.1103/PhysRevPhysEducRes.15.020146

Monroe, C., Raymer, M. G., & Taylor, J. (2019). The US national quantum initiative: from act to action. *Science*, 364(6439), pp.440-442. https://doi.org/10.1126/science.aax0578

Müller, R., & Wiesner, H. (2002). Teaching quantum mechanics on an introductory level. *American Journal of Physics*, 70(3), 200-209. https://doi.org/10.1119/1.1435346

Passante, G., Emigh, J., & Shaffer, P. S. (2015a). Examining student ideas about energy measurements on quantum states across undergraduate and graduate levels. *Physical Review Special Topics - Physics Education Research*, 11(2), 1–10. https://doi.org/10.1103/PhysRevSTPER.11.020111

Passante, G., Emigh, J., & Shaffer, P. S. (2015b). Student ability to distinguish between superposition states and mixed states in quantum mechanics. *Physical Review Special Topics - Physics Education Research*, 11(2), 1–8. https://doi.org/10.1103/PhysRevSTPER.11.020111

Potvin, P., Sauriol, É., & Riopel, M. (2015). Experimental evidence of the superiority of the prevalence model of conceptual change over the classical models and repetition. *Journal of Research in Science Teaching*, 52 (8), pp. 1082-1108. https://doi.org/10.1002/tea.21235

Ravaioli, G., & Levrini, O. (2017). Accepting quantum physics: Analysis of secondary school students' cognitive needs. In O. Finlayson, E. McLoughlin, S. Erduran, & P. Childs (Eds.), *Electronic Proceedings of the ESERA 2017 Conference. Research, Practice and Collaboration in Science Education, Part 2.* https://www.dropbox.com/s/t1ri3ql7ufpihun/Part_2_eBook.pdf?dl=0

Redish, E. F., & Kuo, E. (2015). Language of Physics, Language of Math: Disciplinary Culture and Dynamic Epistemology. *Science & Education*, 24(5-6), 561–590. https://doi.org/10.1007/s11191-015-9749-7

Riedel, M., Kovacs, M., Zoller, P., Mlynek, J., & Calarco, T. (2019). Europe's quantum flagship initiative. *Quantum Science and Technology*, 4, 2, 020501. https://doi.org/10.1088/2058-9565/ab042d

Robertson E., Kohnle A. (2010). Testing the development of student conceptual understanding of quantum mechanics. In D. Raine, C. Hurkett & L. Rogers (Eds.), *Physics community and cooperation: Selected contributions from the GIREP-EPEC & PHEC 2009 International Conference* (pp. 261–273). Lulu / The Centre for Interdisciplinary Science, University of Leicester.

Sandoval, W. A., Green, J. A., & Bråten, I. (2016). Understanding and promoting thinking about knowledge: Origins, issues, and future directions of research on epistemic cognition. *Review of Research in Education*, 40(1), 457-496. https://doi.org/10.3102/0091732X16669319

Sandoval, W. A. (2016). Disciplinary insights into the study of epistemic cognition. In J. A. Greene, W. A. Sandoval, & I. Bråten (Eds.), *Handbook of epistemic cognition* (pp. 184-194). Routledge. https://doi.org/10.4324/9781315795225

Schurtz, G., & Votsis, I. (2014) Reconstructing scientific theory change by means of frames. In T. Gamerschlag, D. Gerland, R. Osswald, R., & W. Petersen (Eds.), *Frames and concept types: Applications in language and philosophy* (pp. 93-109). Springer. https://doi.org/10.1007/978-3-319-01541-5

Sfard, A. (1991). On the dual nature of mathematical conceptions: Reflection on processes and objects as different sides of the same coin. *Educational Studies in Mathematics*, 22(1), 1-36. http://dx.doi.org/10.1007/BF00302715

Schlosshauer, M. (2007). *Decoherence and the quantum-to-classical-transition*. Springer. https://doi.org/10.1007/978-3-540-35775-9

Singh, C. (2001). Student understanding of quantum mechanics. *American Journal of Physics*, 69(8), 885–895. https://doi.org/10.1119/1.1365404

Singh, C. (2007). Helping Students Learn Quantum Mechanics for Quantum Computing. In *Physics Education Research Conference 2006 Proceedings* (pp. 42-45). American Institute of Physics. https://doi.org/10.1063/1.2508687

Singh (2008). Student understanding of quantum mechanics at the beginning of graduate instruction. *American Journal of Physics*, 76(3), 277–287. https://doi.org/10.1119/1.2825387

Singh, C., & Marshman, E. (2015). Review of student difficulties in upper-level quantum mechanics. *Physical Review Special Topics – Physics Education Research*, 11(2), 1-24. https://doi.org/10.1103/PhysRevSTPER.11.020117

Stadermann, H. K. E., van den Berg, E., & Goedhart, M. J. (2019). Analysis of secondary school quantum physics curricula of 15 different countries: Different perspectives on a challenging topic. *Physical Review Physics Education Research*, 15(1), 1-25. https://doi.org/10.1103/PhysRevPhysEducRes.15.010130

Thagard, P. (1992). Conceptual revolutions. Princeton University Press.

Tsaparlis, G. (2013). Learning and teaching the basic quantum chemical concepts. In G. Tsaparlis & H. Sevian (Eds.), *Concepts of matter in science education* (pp. 437-460). Springer. https://doi.org/10.1007/978-94-007-5914-5_21

Uhden, O., Karam, R., Pietrocola, M., & Pospiech, G. (2012). Modelling mathematical reasoning in physics education. *Science & Education*, 21(4), 485–506. https://doi.org/10.1007/s11191-011-9396-6

Von Neumann, J. (1955). *Mathematical foundations of quantum mechanics.* Princeton University Press.

Vosniadou, S. (2008). The framework theory approach to the problem of conceptual change. In S. Vosniadou (Ed.), *International handbook of research on conceptual change* (1st ed.). (pp. 3-34). Routledge. https://doi.org/10.4324/9780203874813

Vosniadou, S., & Mason, L. (2012). Conceptual change induced by instruction: A complex interplay of multiple factors. In K. R. Harris, Steve Graham, T. Urdan, T., Sandra Graham, J. M. Royer, M. & Zeidner (Eds.), *APA* *educational psychology handbook, Vol. 2. Individual differences and cultural and contextual factors* (pp. 221-246). American Psychological Association. https://doi.org/10.1037/13274-009

Vosniadou, S., & Skopeliti, I. (2014). Conceptual change from the framework theory side of the fence. *Science & Education*, 23(7), 1427-1445. https://doi.org/10.1007/s11191-013-9640-3

Vosniadou, S., & Skopeliti, I. (2019). Evaluating the effects of analogy enriched text on the learning of science: The importance of learning indexes. *Journal of Research in Science Teaching*, 56(6), 732-764. https://doi.org/10.1002/tea.21523

Wan, T., Emigh, P. J., & Shaffer, P. S. (2019). Probing student reasoning in relating relative phase and quantum phenomena. *Physical Review Physics Education Research*, 15(2), 1-13. https://doi.org/10.1103/PhysRevPhysEducRes.15.020139

Wittmann, M. C., Morgan, J. T, Bao, L. (2005). Addressing student models of energy loss in quantum tunnelling. *European Journal of Physics,* 26(6), 939-950. https://doi.org/10.1088/0143-0807/26/6/001

Zenker, F. (2014). From features via frames to spaces. Modeling scientific conceptual change without incommensurability or aprioricity. In T. Gamerschlag, D. Gerland, R. Osswald, & W. Petersen (Eds.), *Frames and concept types: Applications in language and philosophy* (pp. 69–89). Springer. https://doi.org/10.1007/978-3-319-01541-5

Zohar, A. R., & Levy, S. T. (2021). From feeling forces to understanding forces: The impact of bodily engagement on learning in science. *Journal of Research in Science Teaching*, 58(5), 1-35. https://doi.org/10.1002/tea.21698

Zhu, G., & Singh, C. (2012). Surveying students' understanding of quantum mechanics in one spatial dimension. *American Journal of Physics*, 80(3), 252-259. https://doi.org/10.1119/1.3677653